\documentclass{article}

 \usepackage[eandd,final,nonatbib]{neurips_2026}

\usepackage{amsmath,graphicx}
\usepackage{iftex}
\usepackage{fontspec}
\usepackage{newunicodechar}
\newfontfamily\akanfont{NotoSerif-Regular.ttf}[Path=fonts/,Scale=MatchLowercase]
\newunicodechar{ɔ}{{\akanfont ɔ}}
\newunicodechar{ɛ}{{\akanfont ɛ}}
\newunicodechar{Ɛ}{{\akanfont Ɛ}}
\newfontfamily\arabicfont{NotoSansArabic-Regular.ttf}[Path=fonts/,Script=Arabic,AutoFakeSlant=0.15]
\newfontfamily\ethiopicfont{NotoSansEthiopic-Regular.ttf}[Path=fonts/,Script=Ethiopic,AutoFakeSlant=0.15]
\newfontfamily\sinhalafont{NotoSansSinhala-Regular.ttf}[Path=fonts/,Script=Sinhala]
\newfontfamily\cyrillicfont{NotoSans-Regular.ttf}[Path=fonts/,Script=Cyrillic]
\ifXeTeX
  \newcommand{\textarabic}[1]{{\beginR\arabicfont #1\endR}}
\else
  \newcommand{\textarabic}[1]{{\textdirection 1\arabicfont #1}}
\fi
\newcommand{\textethi}[1]{{\ethiopicfont #1}}
\newcommand{\textsinhala}[1]{{\sinhalafont #1}}
\newcommand{\textcyrillic}[1]{{\cyrillicfont #1}}
\usepackage{xltabular}
\usepackage{makecell}
\newfontfamily\telugufont{NotoSansTelugu-Regular.ttf}[Path=fonts/,Script=Telugu]
\newcommand{\texttelugu}[1]{{\telugufont #1}}
\newfontfamily\hebrewfont{NotoSansHebrew-Regular.ttf}[Path=fonts/,Script=Hebrew]
\newcommand{\texthebrew}[1]{{\hebrewfont #1}}
\newfontfamily\devanagarifont{NotoSansDevanagari-Regular.ttf}[Path=fonts/,Script=Devanagari]
\newcommand{\textdevanagari}[1]{{\devanagarifont #1}}
\newfontfamily\greekfont{NotoSans-Regular.ttf}[Path=fonts/,Script=Greek]
\newcommand{\textgreek}[1]{{\greekfont #1}}
\newfontfamily\thaanafont{NotoSansThaana-Regular.ttf}[Path=fonts/,Script=Thaana]
\newcommand{\textthaana}[1]{{\thaanafont #1}}
\newfontfamily\bengalifont{NotoSansBengali-Regular.ttf}[Path=fonts/,Script=Bengali]
\newcommand{\textbengali}[1]{{\bengalifont #1}}
\newfontfamily\malayalamfont{NotoSansMalayalam-Regular.ttf}[Path=fonts/,Script=Malayalam]
\newcommand{\textmalayalam}[1]{{\malayalamfont #1}}
\newfontfamily\tibetanfont{NotoSerifTibetan-Regular.ttf}[Path=fonts/,Script=Tibetan]
\newcommand{\texttibetan}[1]{{\tibetanfont #1}}

\usepackage{url}            %
\usepackage{booktabs}       %
\usepackage{amsfonts}       %
\usepackage{nicefrac}       %
\usepackage{microtype}      %
\usepackage{xcolor}         %
\usepackage{multirow}

\usepackage{enumitem}
\usepackage{afterpage}
\definecolor{darkblue}{RGB}{0, 0, 130}

\usepackage[
    colorlinks=true,
    linkcolor=magenta,
    citecolor=darkblue,
    urlcolor=blue
]{hyperref}                 %

\title{WAXAL: A Large-Scale Multilingual African Language Speech Corpus}

\author{
\mdseries
\textit{Abdoulaye Diack}$^1$,
\textit{Perry Nelson}$^1$,
\textit{MohamedElfatih MohamedKhair}$^1$,
\textit{Emmanuel Asiedu Brempong}$^1$\\
\textit{Subhashini Venugopalan}$^1$,
\textit{Tavonga Siyavora}$^1$,
\textit{Vusumuzi Dube}$^1$,
\textit{Kwaku Agbesi}$^1$,
\textit{Angela Nakalembe}$^1$,\\
\textit{Jason Hickey}$^1$,
\textit{Uche Okonkwo}$^1$,
\textit{Abhishek Bapna}$^1$,
\textit{Isaac Wiafe}$^2$,
\textit{Raynard Dodzi Helegah}$^2$,\\
\textit{Elikem Doe Atsakpo}$^2$,
\textit{Charles Nutrokpor}$^2$,
\textit{Fiifi Baffoe Payin Winful}$^2$,
\textit{Kafui Kwashie Solaga}$^2$,\\
\textit{Jamal-Deen Abdulai}$^2$,
\textit{Akon Obu Ekpezu}$^2$,
\textit{Audace Niyonkuru}$^3$,
\textit{Samuel Rutunda}$^3$,
\textit{Boris Ishimwe}$^3$,\\
\textit{Michael Melese}$^4$,
\textit{Engineer Bainomugisha}$^5$,
\textit{Joyce Nakatumba-Nabende}$^5$,
\textit{Andrew Katumba}$^5$,\\
\textit{Claire Babirye}$^5$,
\textit{Jonathan Mukiibi}$^5$,
\textit{Vincent Kimani}$^6$,
\textit{Samuel Kibacia}$^6$,
\textit{James Maina}$^6$,
\textit{Fridah Emmah}$^6$,\\
\textit{Ahmed Ibrahim Shekarau}$^7$,
\textit{Ibrahim Shehu Adamu}$^7$,
\textit{Yusuf Abdullahi}$^7$,
\textit{Howard Lakougna}$^8$,
\textit{Kode Niane}$^9$,\\
\textit{Bob MacDonald}$^1$,
\textit{Pooja Rao}$^1$,
\textit{Hadar Shemtov}$^1$,
\textit{Prince Mireku}$^1$,\\
\textit{Aisha Walcott-Bryant}$^1$,
\textit{Moustapha Cisse},
\textit{Avinatan Hassidim}$^1$,
\textit{Jeff Dean}$^1$,
\textit{Yossi Matias}$^1$\\
$^1$Google Research,
$^2$University of Ghana,
$^3$Digital Umuganda,
$^4$Addis Ababa University,
$^5$Makerere\\ University,
$^6$Loud and Clear Comm. Ltd.,
$^7$Media Trust Ltd.,
$^8$Gates Foundation,
$^9$AIMS Senegal
}

\begin{document}
\vspace{-0.4cm}
\maketitle

\vspace{-0.2cm}
\begin{abstract}
\vspace{-0.2cm}
Recent advances in speech technology predominantly favor high-resource languages, leaving speakers of Sub-Saharan African languages underserved. To bridge this divide, we introduce WAXAL, an open, large-scale speech dataset spanning 27 African languages representing over 100 million speakers. WAXAL comprises over 14,100 hours of speech collected in partnership with African academic and community organizations to build local research capacity: an Automatic Speech Recognition (ASR) corpus with 2,340 hours of transcribed, image-prompted natural speech across 19 languages (plus $\sim$11,760 untranscribed hours) and a studio-quality Text-to-Speech (TTS) corpus of $\sim$247 hours across 15 language configurations. We empirically benchmark three architecturally distinct ASR models---Gemma 3n 2B, Whisper-Large-v3, and MMS-1B-all---across the 19 ASR languages on strictly speaker-disjoint splits. Fine-tuning on WAXAL yields substantial gains, reducing macro-average WER by up to 58\% for Whisper-Large-v3, 56\% for Gemma 3n 2B, and 17\% for MMS-1B-all (and Character Error Rate by up to 70\%). WAXAL-fine-tuned models generalize zero-shot to three out-of-distribution benchmarks: FLEURS~\cite{conneau2023fleurs}, Mozilla Common Voice~\cite{ardila2019common}, and Sunbird SALT~\cite{akera2022salt}. Finally, an LLM-as-judge semantic evaluation confirms that fine-tuning preserves speaker intent beyond surface-level transcription metrics. WAXAL is released at \href{https://huggingface.co/datasets/google/WaxalNLP}{hf.co/datasets/google/WaxalNLP}.

\end{abstract}
\vspace{-0.2cm}
\section{Introduction}
\label{sec:intro}
\vspace{-0.1cm}

The proliferation of voice-enabled technologies, from virtual assistants to automated transcription services, has transformed human-computer interaction. However, the benefits of these advancements are unevenly distributed, with a stark concentration in a handful of high-resource languages. This has left hundreds of millions of people, particularly in Sub-Saharan Africa, unable to access technologies in their native tongues. The primary bottleneck for developing robust Automatic Speech Recognition (ASR) and Text-to-Speech (TTS) systems is the profound scarcity of large-scale, high-quality, and permissively licensed speech corpora.

\begin{figure*}[!htb]
\vspace{-0.2cm}

\begin{minipage}[b]{.245\linewidth}
  \centering
  \centerline{\includegraphics[width=4.4cm,height=3cm]{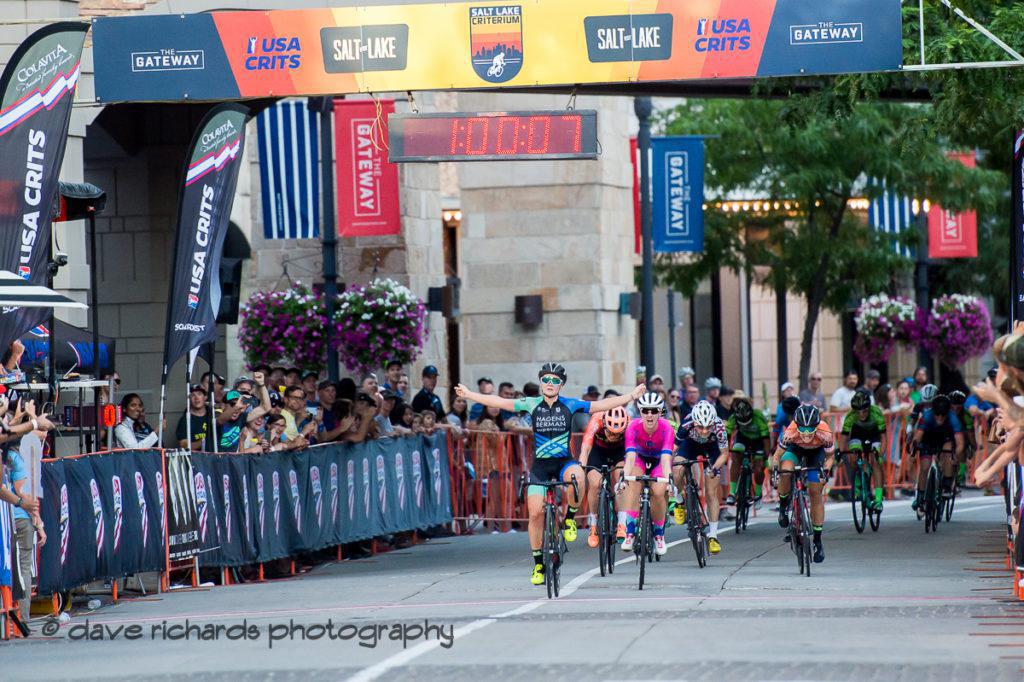}}
\end{minipage}
\begin{minipage}[b]{0.245\linewidth}
  \centering
  \centerline{\includegraphics[width=4.4cm,height=3cm]{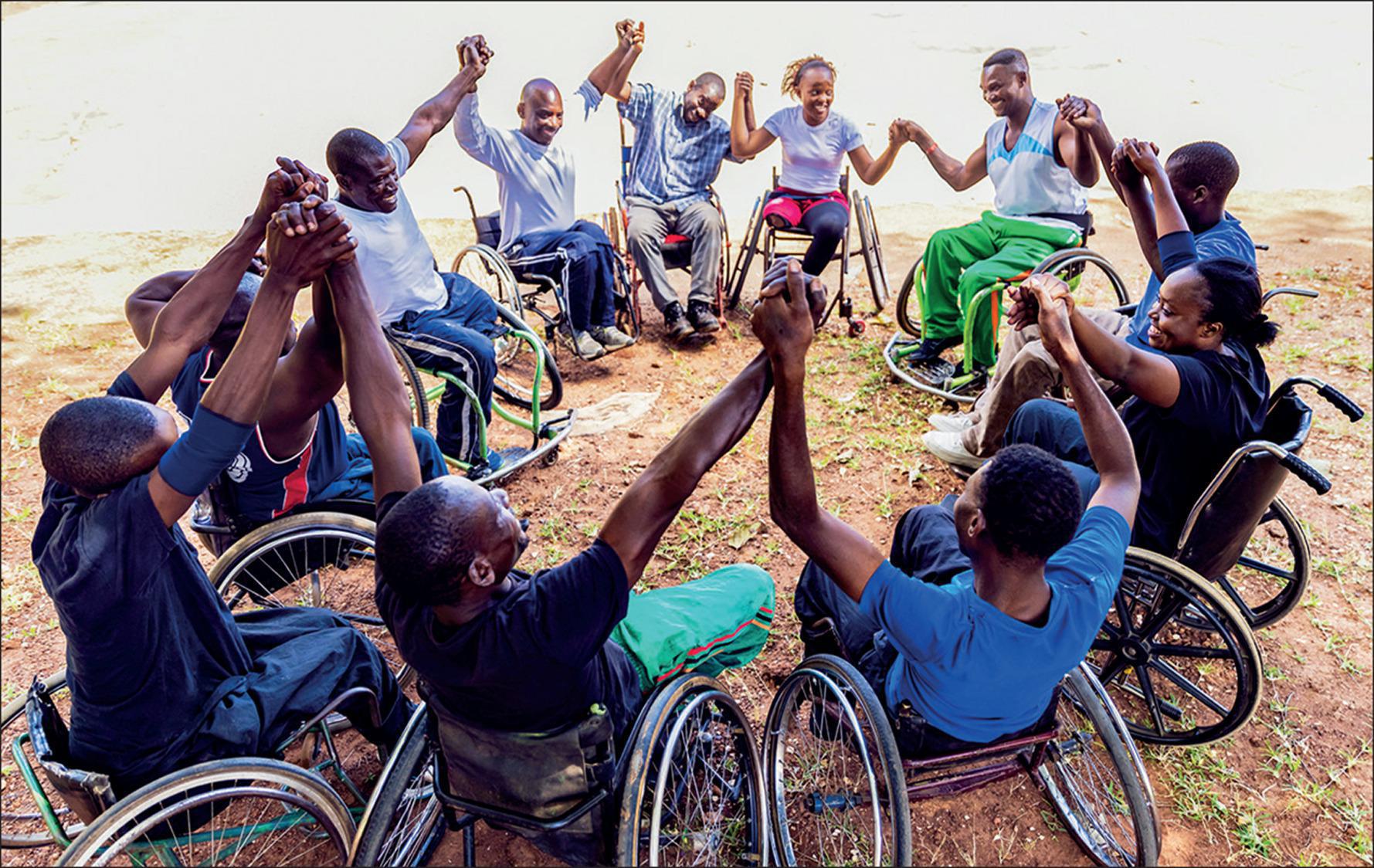}}
\end{minipage}
\begin{minipage}[b]{.245\linewidth}
  \centering
  \centerline{\includegraphics[width=4.4cm,height=3cm]{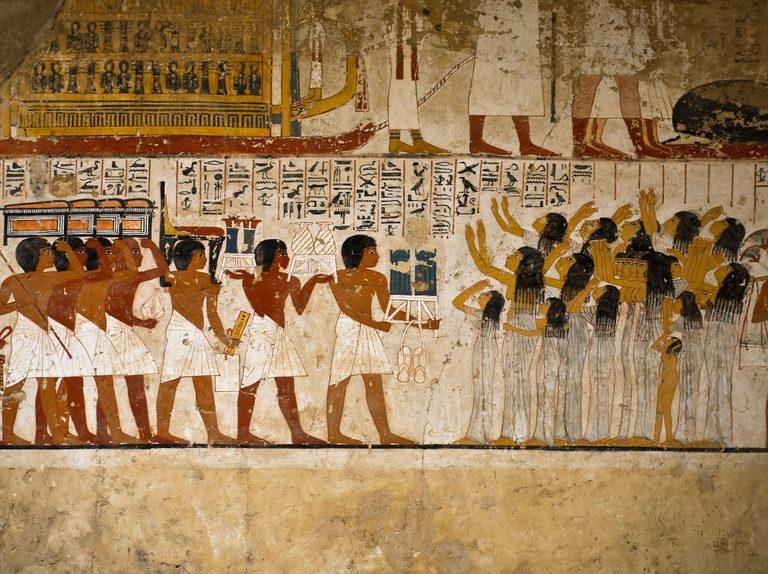}}
\end{minipage}
\begin{minipage}[b]{.245\linewidth}
  \centering
  \centerline{\includegraphics[width=4.4cm,height=3cm]{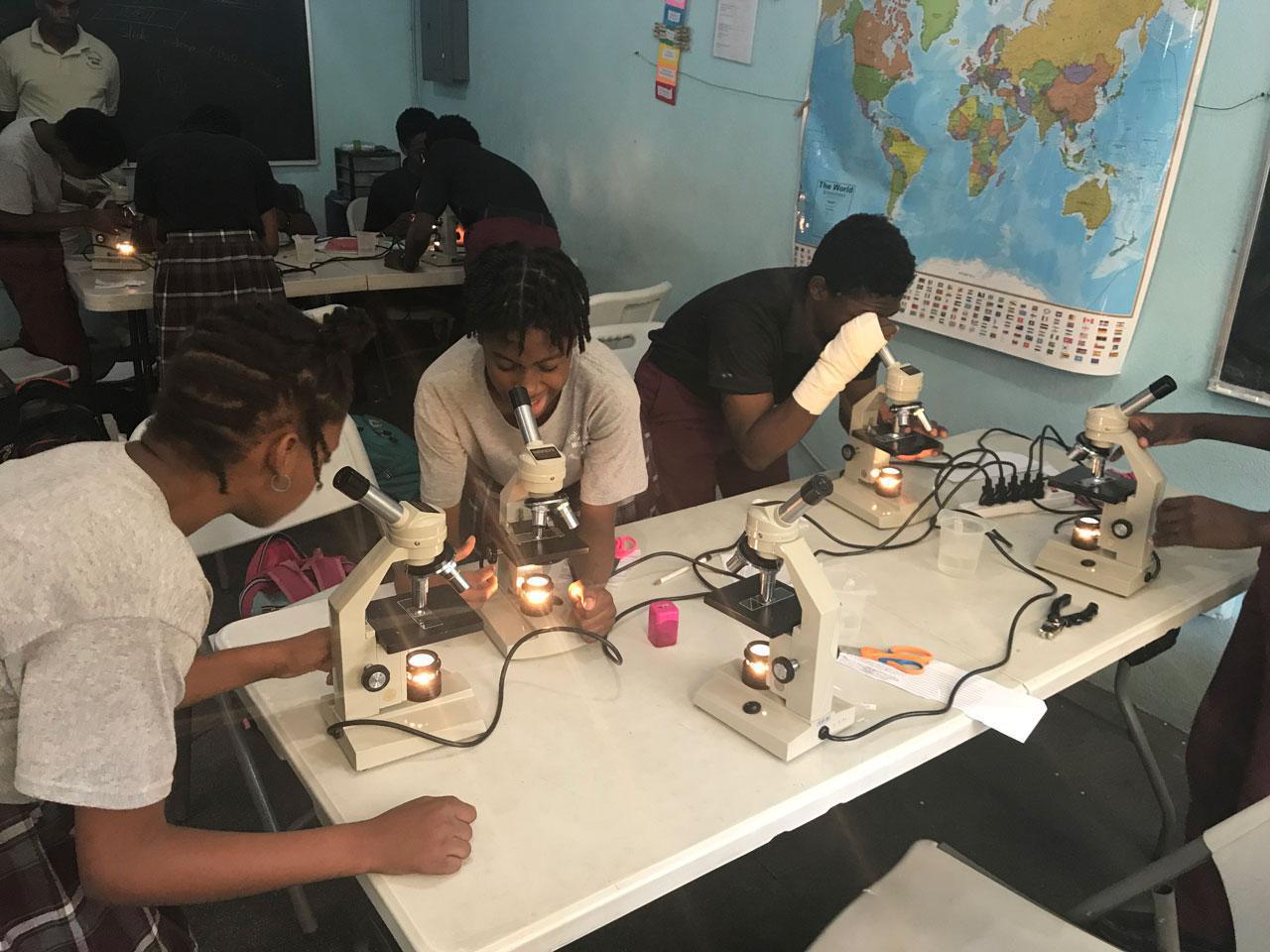}}
\end{minipage}
\begin{minipage}[b]{.245\linewidth}
  \centering
  \centerline{\includegraphics[width=4.4cm,height=3cm]{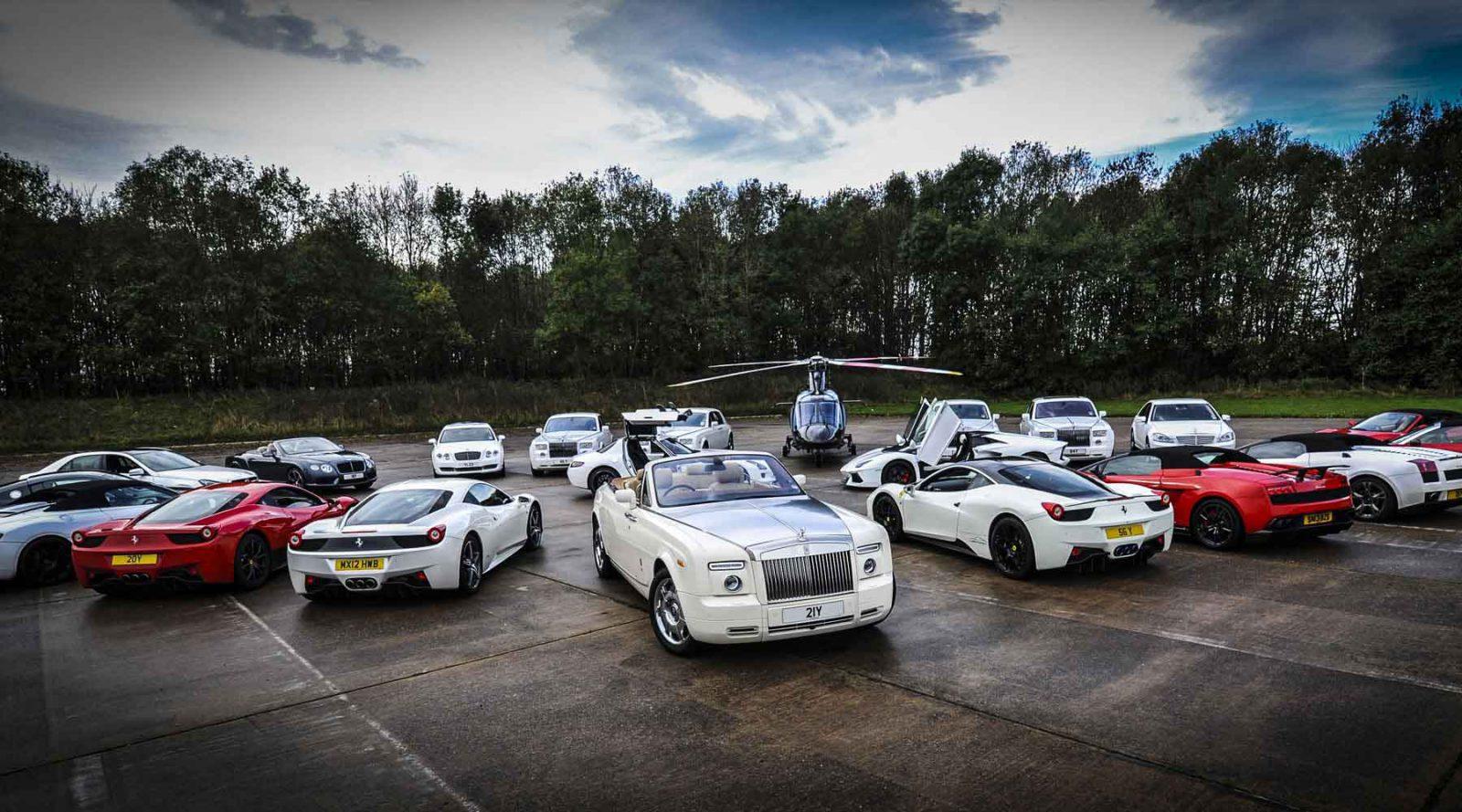}}
\end{minipage}
\begin{minipage}[b]{.245\linewidth}
  \centering
  \centerline{\includegraphics[width=4.4cm,height=3cm]{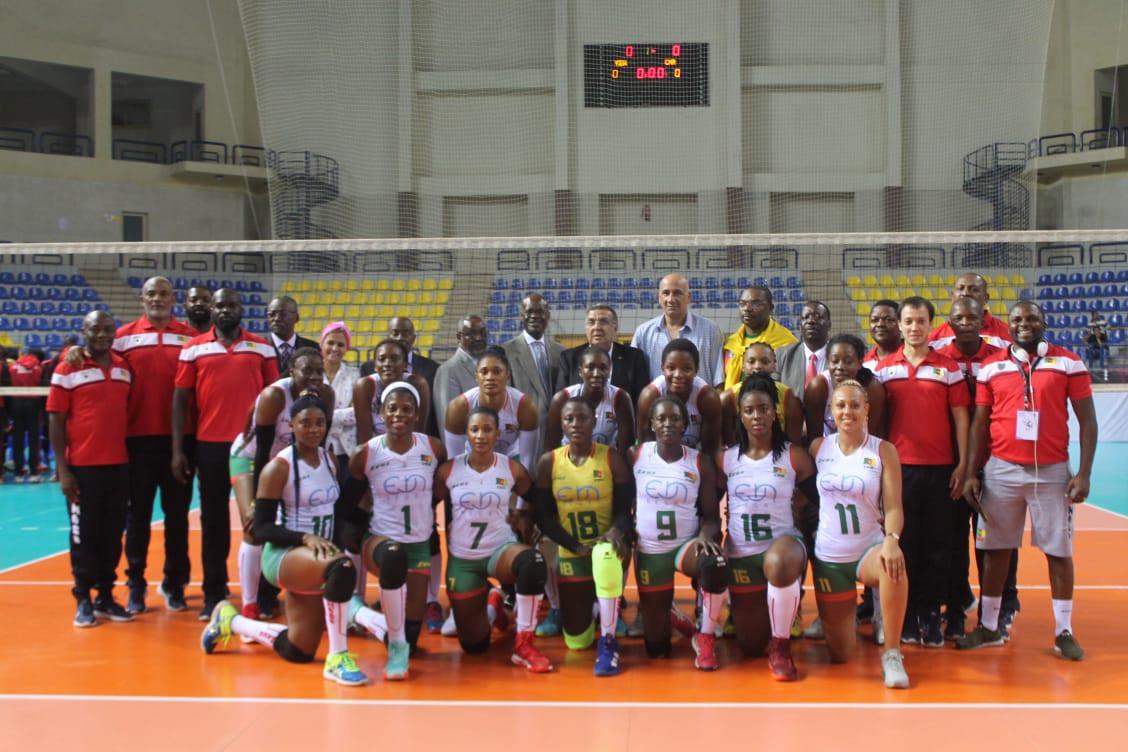}}
\end{minipage}
\begin{minipage}[b]{.245\linewidth}
  \centering
  \centerline{\includegraphics[width=4.4cm,height=3cm]{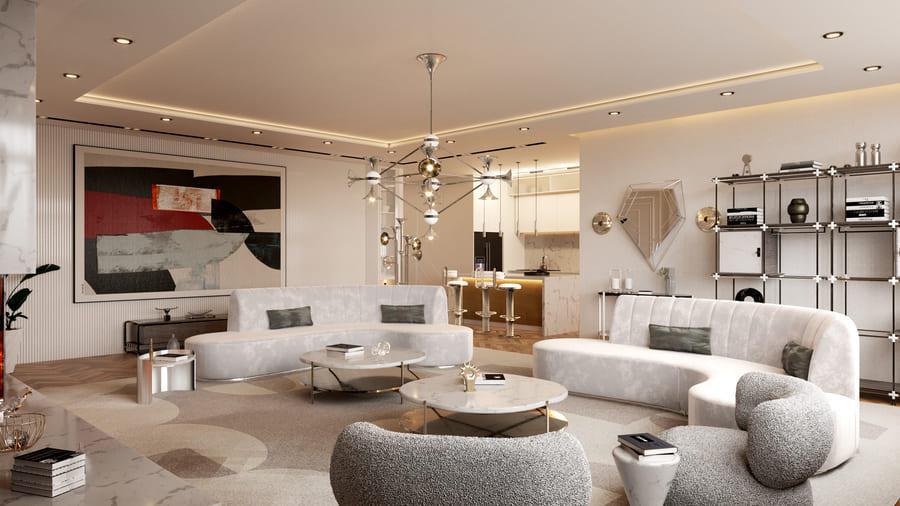}}
\end{minipage}
\begin{minipage}[b]{.245\linewidth}
  \centering
  \centerline{\includegraphics[width=4.4cm,height=3cm]{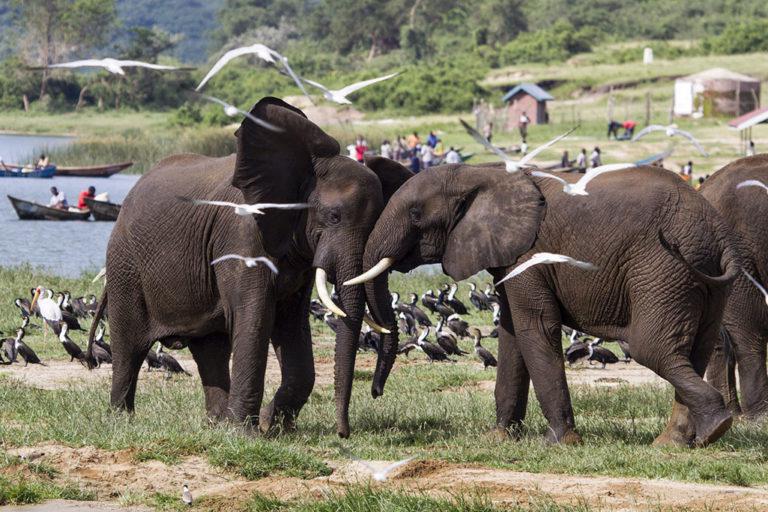}}
\end{minipage}
\caption{Examples of some of the images used as prompts to elicit natural speech for the ASR dataset.}
\label{fig:image_prompt}
\vspace{-0.4cm}
\end{figure*}

Africa, home to 1.3 billion people and over 2,000 distinct languages, represents one of the most linguistically diverse regions on the planet. Despite this richness, Sub-Saharan Africa remains critically underserved by modern speech technology largely due to the scarcity of large-scale speech corpora in many of these languages.
This data deficit is exacerbated by the unique linguistic characteristics of the region, such as tonal distinctions, complex morphology, and prevalent code-switching, which pose significant challenges for standard modeling paradigms. Without foundational datasets, it is impossible to train, evaluate, and adapt models to serve the educational, accessibility, and communication needs of these communities.
To address this critical need, we present the WAXAL dataset, a new, large-scale resource for 27 Sub-Saharan African languages (comprising a robust ASR benchmarking corpus spanning 19 languages and a complementary TTS data release across 15 language configurations, with speaker-diversity verification and TTS synthesis ablations in Appendix~\ref{app:tts_evaluation}) collected in partnership with five African academic and community partners, with the goal of building the local AI ecosystem and seeding data collection capabilities. %
The collection features both ASR and TTS data for languages such as Acholi, Luganda, Akan, and Ewe; ASR data for Shona, Fulani, and Lingala; and TTS data for Kiswahili, Hausa, Igbo, and Yoruba, among others.
Our contributions are:
\begin{itemize}[noitemsep,nolistsep,leftmargin=8pt]
    \item \textbf{A Large-Scale ASR Corpus:} We release $\sim$2,340 hours of transcribed, image-prompted natural speech from 5,021 speakers, alongside $\sim$11,760 hours of untranscribed audio across 19 languages, suitable for self-supervised learning, training, and speaker-disjoint evaluation of ASR models.
    \item \textbf{A High-Quality TTS Corpus:} We provide $\sim$247 hours ($\sim$25k utterances, 103 speakers) of studio-quality recordings of phonetically balanced scripts across 15 languages. %
    \item \textbf{Rich Metadata and Quality Assurance:} The data is accompanied by speaker demographics and recording environment metadata, transcribed by local language experts with double-annotator verification (achieving 95.2\% character-level agreement; 4.8\% inter-annotator CER).
    \item \textbf{Comprehensive ASR Benchmarks:} We evaluate three architecturally distinct models---Gemma 3n 2B, Whisper-Large-v3, and MMS-1B-all---and show that fine-tuning on WAXAL substantially reduces macro-average WER: by 58\% for Whisper-Large-v3, 56\% for Gemma 3n 2B, and 17\% for MMS-1B-all. Further, these fine-tuned models transfer effectively zero-shot to three out-of-distribution benchmarks (FLEURS~\cite{conneau2023fleurs}, Mozilla Common Voice~\cite{ardila2019common}, and Sunbird SALT~\cite{akera2022salt}).
    \item \textbf{Semantic Evaluation via LLM-as-Judge:} Beyond surface-level metrics like WER and CER, we introduce an LLM-as-judge evaluation to assess semantic meaning preservation, validated against native human judgments (achieving Pearson $r = 0.71$, $\text{MAE} = 0.14$).
\end{itemize}

\vspace{-0.3cm}
\section{Related Works}
\label{sec:related_works}
In recent years, several significant multilingual datasets have been released, aiming to broaden linguistic diversity. Prominent examples include Common Voice \cite{ardila2019common}, which has crowdsourced over 19k hours of recordings across 290 languages, and FLEURS \cite{conneau2023fleurs}, which covers 102 languages with approximately 12 hours of supervised data each. Other pioneering efforts like the BABEL project provided conversational telephone speech for 17 low-resource languages \cite{gales2014speech}, while the CMU Wilderness dataset \cite{black2019cmu} contains recordings in over 700 languages. For unsupervised and semi-supervised applications, datasets like VoxLingua107 \cite{valk2021voxlingua107} and YODAS~\cite{li2023yodas} offer thousands of hours of unlabeled speech from YouTube. However, a significant portion of these resources is concentrated on high-resource languages, and even those targeting low-resource settings, such as GigaSpeech~2 \cite{yang2024gigaspeech}, still lack substantial representation of African languages.

When focusing specifically on Sub-Saharan African languages, existing public datasets are often limited in scale, scope, or speaker diversity.
For ASR, prior work includes datasets derived from radio archives, such as \cite{doumbouya2021usingradio} containing $\sim$10k utterances from 49 speakers. The ALFFA project made valuable contributions by collecting prompted speech for four languages, including Wolof \cite{gauthier2016collecting}.
More recently, Afri-voice~\cite{niyonkuru2024afrivoice} released an image-prompted and scripted corpus covering six African languages---Shona, Lingala, Fulani, Malagasy, Wolof, and Somali ($\sim$600 transcribed hours total)---following a collection methodology similar to ours.
Afrivox~\cite{intronhealth2024afrivox} provides $\sim$64 hours across 18 African languages in both medical and non-medical domains, serving as a benchmark for ASR and speech translation.
To support evaluation and comparison across African-language ASR systems, PazaBench~\cite{microsoft2024pazabench} aggregates existing datasets and offers a public leaderboard specifically targeting low-resource African languages.
Similarly, resources for TTS in African languages remain scarce. \cite{van-niekerk-etal-2017} detailed a methodology for rapidly developing high-quality corpora for four South African languages. More recently, BibleTTS \cite{meyer2022bibletts} released a significant corpus containing over 80 hours of single-speaker, studio-quality aligned speech for six languages spoken in Sub-Saharan Africa. While these datasets are invaluable, they cover a limited number of languages and speakers.

\begin{figure*}[!tb]
\vspace{-0.5cm}
\begin{minipage}[b]{0.6\linewidth}
  \centering
  \centerline{\includegraphics[height=6cm]{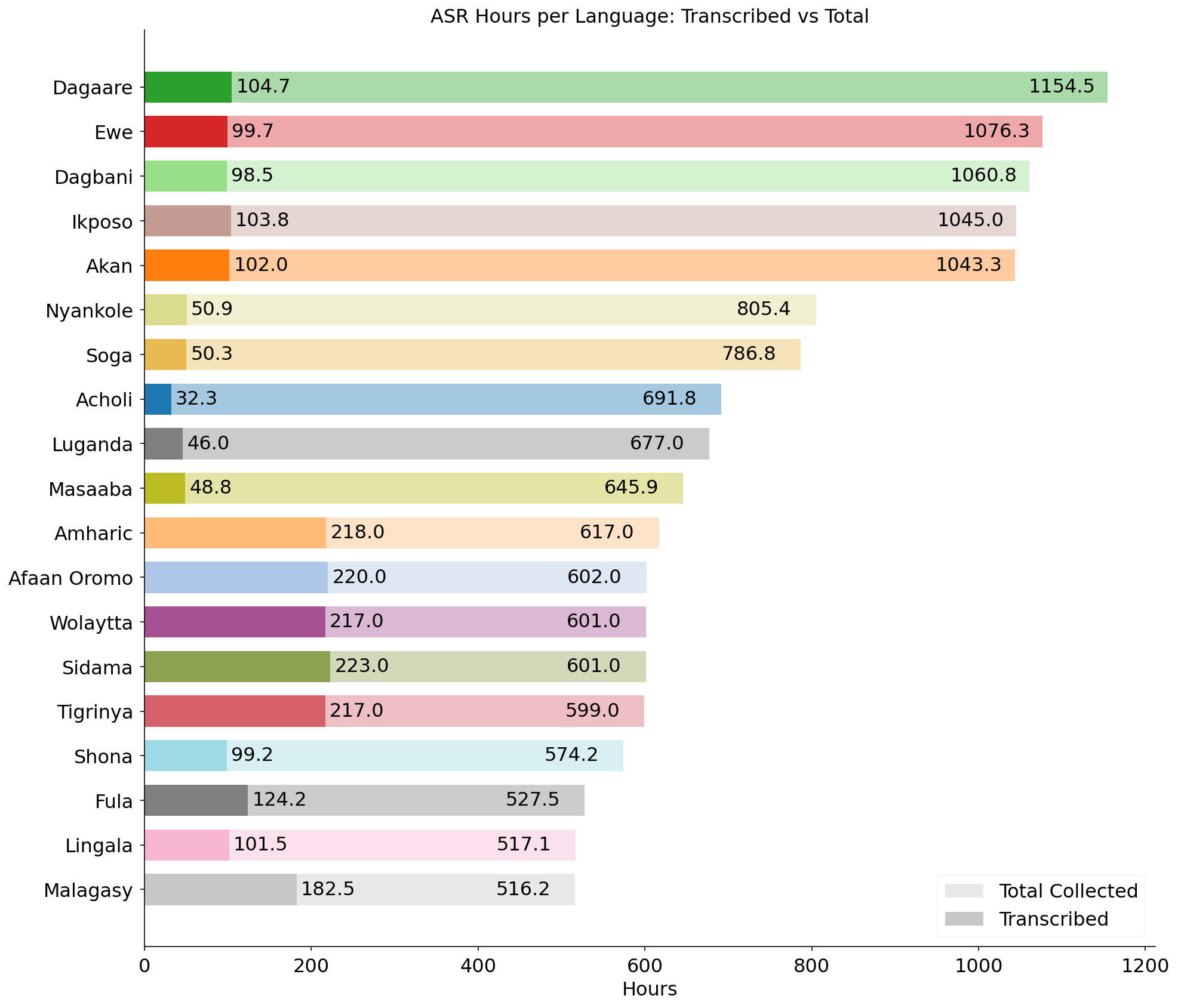}}
  \centerline{(a) WAXAL-ASR: Transcribed hrs and Total hrs per language.}\medskip
\vspace{-0.4cm}
\end{minipage}
\hfill
\begin{minipage}[b]{.4\linewidth}
  \centering
  \centerline{\includegraphics[width=\linewidth]{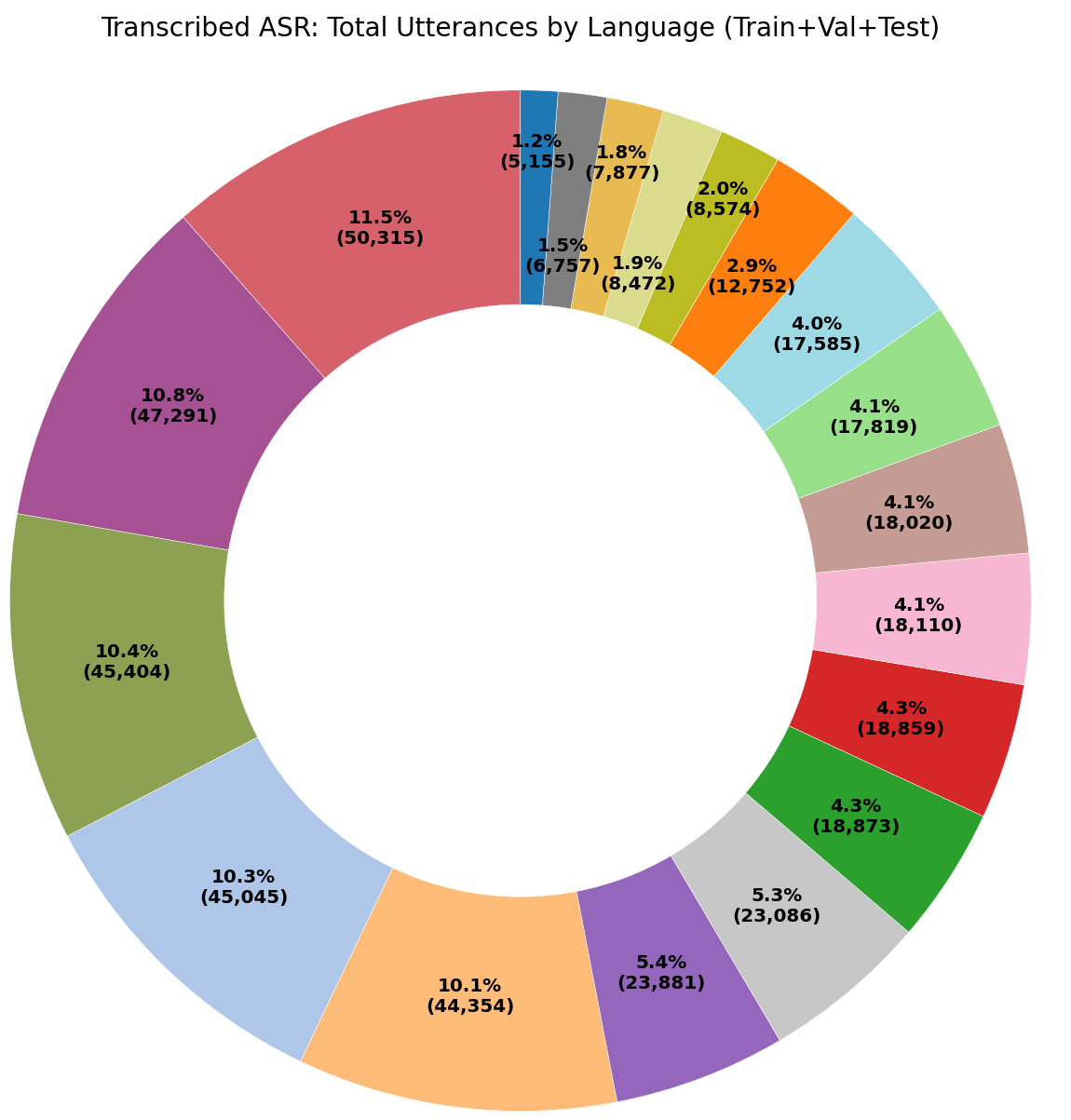}}
  \centerline{(b) Transcribed Utterances per language}\medskip
\vspace{-0.4cm}
\end{minipage}
\caption{WAXAL-ASR: Distribution of the transcribed and total recordings per language.}
\label{fig:asr_trans_tts_utts_dist}
\end{figure*}

Our work directly addresses these gaps by introducing a large-scale, unified dataset for 27 Sub-Saharan African languages. While massive datasets like MMS~\cite{pratap2024scaling} cover many languages, their data distribution is highly \textit{skewed}, meaning a few languages dominate while most have very limited coverage. Furthermore, MMS relies predominantly on scripted reading (often religious texts), lacking the spontaneous and natural speech characteristics found in everyday communication. In contrast, WAXAL provides a more balanced distribution (ranging from 30 to 220 hours per language) of image-prompted, spontaneous speech, making it more representative of natural language use. Table~\ref{tab:dataset_comparison} provides a direct comparison against related datasets, highlighting WAXAL's unique combination of scale, language focus, and spontaneous elicitation.

\begin{table}[t]
\centering
\footnotesize
\setlength{\tabcolsep}{3pt}
\caption{WAXAL has spontaneous elicitation and uniform scale vs. related African speech datasets.}
\label{tab:dataset_comparison}
\begin{tabular}{lcccccc}
\toprule
\textbf{Dataset} & \textbf{\# Langs} & \textbf{\# African} & \textbf{ASR Hrs} & \textbf{Avg Hrs/Lang} & \textbf{Elicitation} & \textbf{Domain} \\
\midrule
FLEURS~\cite{conneau2023fleurs}         & 102   & $\sim$20 & $\sim$1,200   & $\sim$12      & Scripted reading           & Wikipedia/News \\
Common Voice~\cite{ardila2019common}   & 290   & 80+      & 19,000+       & Skewed        & Crowdsourced reading       & Open/Varied \\
MMS-Lab~\cite{pratap2024scaling}            & 1,107 & 300+     & 500,000+      & Skewed        & Scripted reading           & Religious text \\
Afri-voice~\cite{niyonkuru2024afrivoice} & 6     & 6        & 600           & 100           & Image-prompted \& Scripted & Open-domain \\
Afrivox~\cite{intronhealth2024afrivox}  & 18    & 16       & $\sim$64      & 0.01--6       & ---                        & Medical \\
\textbf{WAXAL (ours)}                   & \textbf{27}   & \textbf{27} & \textbf{$\sim$2,340} & \textbf{$\sim$30--220} & \textbf{Image-Prompted, Spontaneous} & \textbf{Open/Varied} \\
\bottomrule
\end{tabular}
\vspace{-0.5cm}
\end{table}

\vspace{-0.3cm}
\section{Data Collection Methodology}
\vspace{-0.2cm}
\label{sec:data_collection}

WAXAL was curated (Jan 2021--Mar 2024) with five African partners to ensure local expertise and community involvement: \textbf{Makerere University} (5 ASR, 4 TTS languages), \textbf{University of Ghana} (5 ASR languages; 4 TTS configurations: Twi and Fante varieties of Akan, Baoul\'e, and Ewe), \textbf{Digital Umuganda} (9 ASR languages), \textbf{Media Trust} (5 TTS languages), and \textbf{Loud and Clear} (TTS for Kikuyu and Luo, plus supplementary Luganda and Kiswahili).

Crucial to the development of this resource was a collaborative
framework designed for capacity building within the African AI
ecosystem. While our team provided technical support and guidance,
the collection efforts were led entirely by our African academic
and community partners. A foundational principle of this
relationship was that partners retained full ownership of the
collected data. This partner-led approach ensured the corpus was
built by and for the community it serves, empowering local
institutions to drive derivative research while maintaining a
shared commitment to making the dataset openly accessible to the
broader public.

\begin{figure*}[!tb]
\vspace{-0.3cm}

\begin{minipage}[b]{0.47\linewidth}
  \centering
  \centerline{\includegraphics[width=\linewidth]{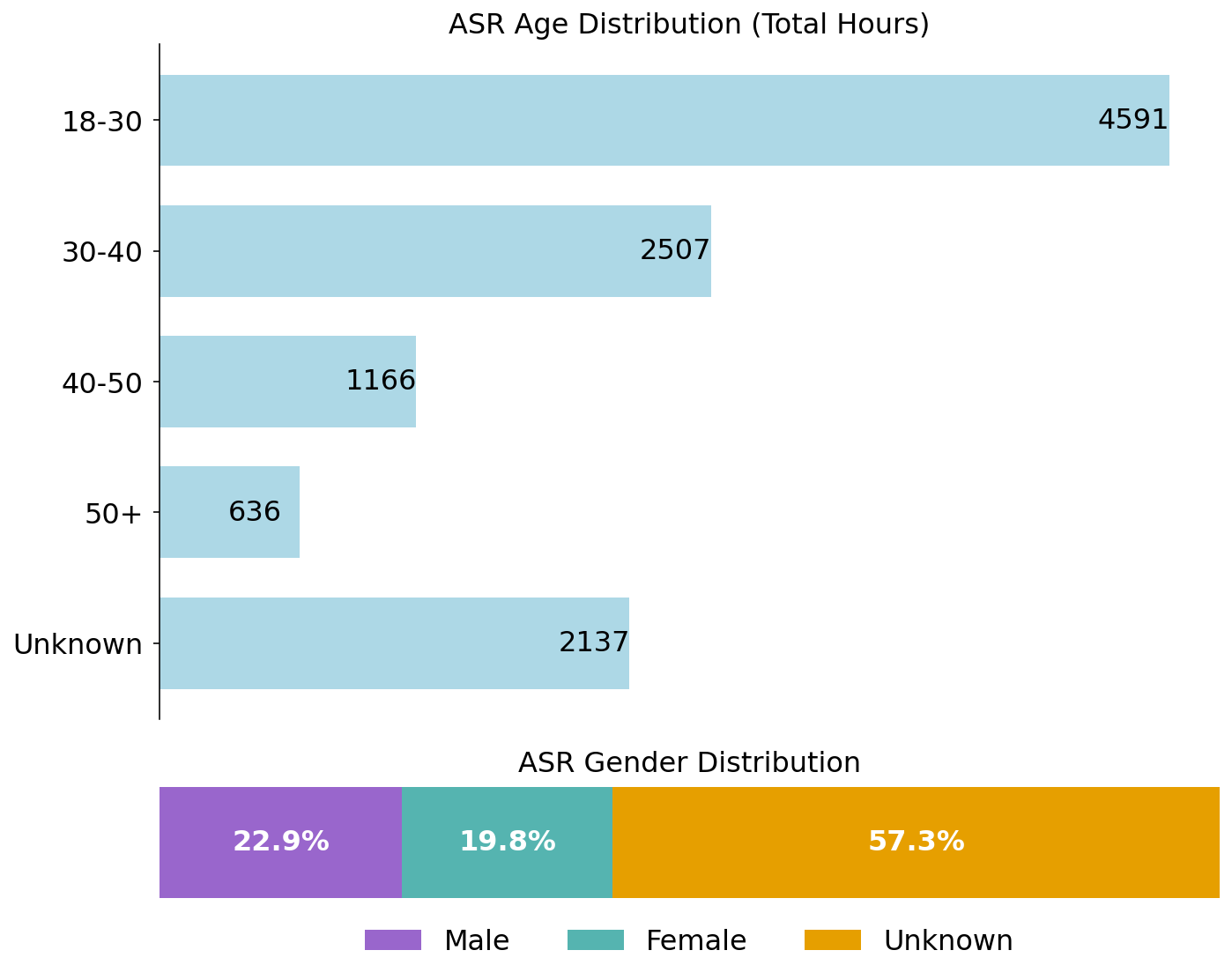}}
  \centerline{(a) Demographics of ASR participants.}\medskip
  \vspace{-0.3cm}
\end{minipage}
\hfill
\begin{minipage}[b]{.49\linewidth}
  \centering
  \centerline{\includegraphics[width=\linewidth]{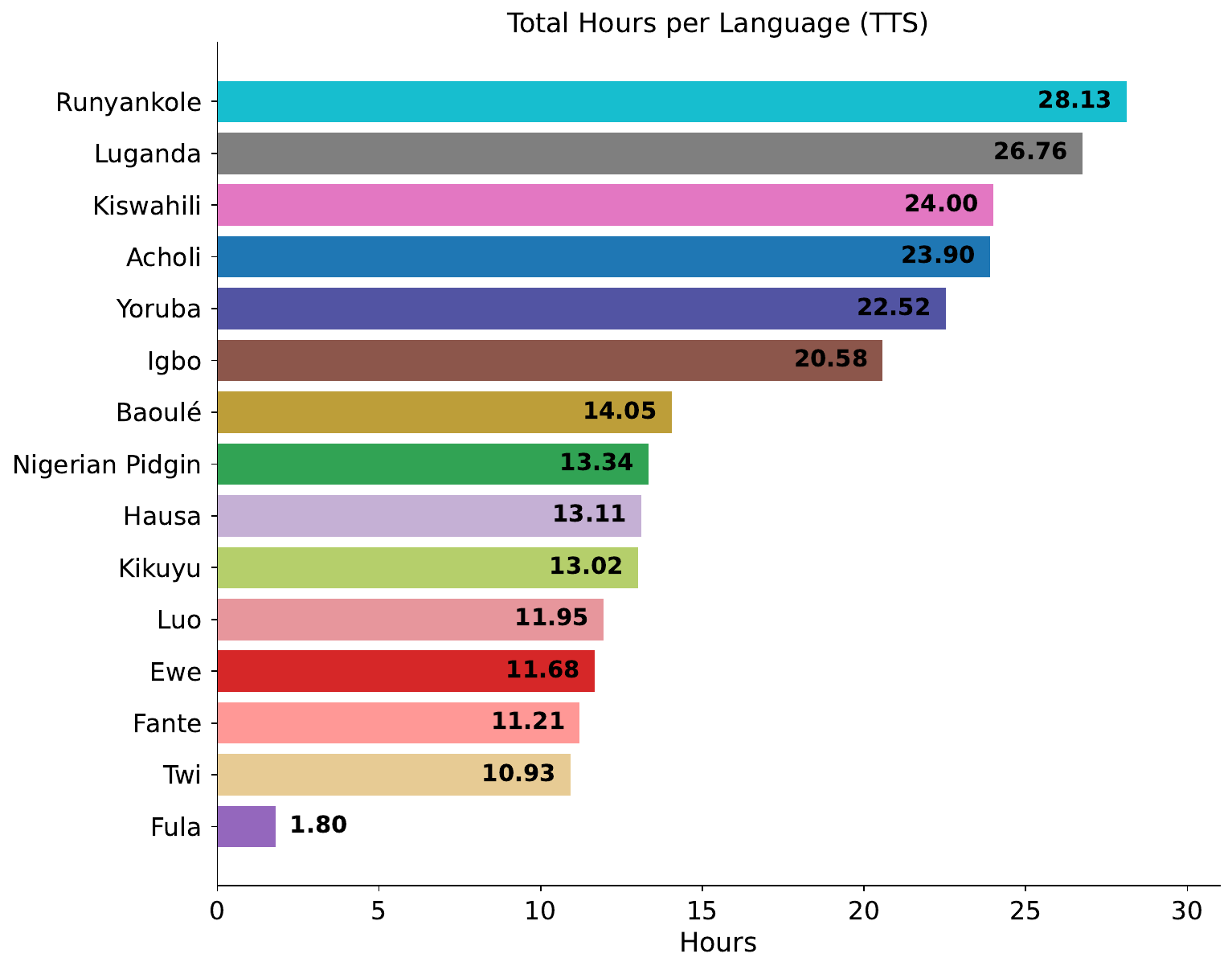}}
  \centerline{(b) WAXAL-TTS: Hours of recordings.}\medskip
  \vspace{-0.3cm}
\end{minipage}

\caption{Demographics of ASR participants and Per-Language Distribution of TTS data.}
\label{fig:asr_demo_tts_dist}
\vspace{-0.5cm}
\end{figure*}

\vspace{-0.2cm}
\subsection{ASR Data Collection and Annotation}
\vspace{-0.2cm}
\label{subsec:asr_data_coll}
The ASR dataset was designed to capture natural, spontaneous speech across four major language phyla spoken in Sub-Saharan Africa and Madagascar: \textbf{Niger-Congo} (Akan, Dagaare, Dagbani, Ewe, Fula, Ikposo, Lingala, Luganda, Masaaba, Runyankole, Shona, Soga), \textbf{Nilo-Saharan} (Acholi), \textbf{Afroasiatic} (Semitic: Amharic and Tigrinya written in the Ge'ez \textit{abugida} script; Cushitic: Oromo and Sidama; Omotic: Wolaytta), and \textbf{Austronesian} (Malagasy). This typological breadth encompasses highly agglutinative Bantu and Atlantic verbal morphology as well as lexical and grammatical tone systems. \textbf{Image-Prompted Speech} involved showing participants a diverse set of images covering at least 50 culturally grounded topics and asking them to describe them in their native language to elicit natural speech~\cite{vaani2025}. \textbf{Recordings} were captured in the speaker's natural environment (minimum 15 seconds, standardized to 16\,kHz 16-bit PCM audio) aiming for gender and age balance across 5,021 contributors. \textbf{Transcription} of approximately 10--17\% of the audio was performed by paid, local linguistic experts using standard local orthographies. The remaining $\sim$11,760 untranscribed hours are fully released alongside the annotated dataset to support self-supervised and semi-supervised learning. Finally, rigorous \textbf{Quality Control} ensured acoustic clarity, prompt relevance, and PII removal: $5.1\%$ of raw audio segments were rejected during validation due to excessive background noise, truncation, or prompt mismatch, and over $30\%$ of transcribed utterances underwent independent double-annotator verification, achieving an average inter-annotator Character Error Rate (CER) of $4.8\%$ ($95.2\%$ character-level agreement).
The intentionally collected attributes for the ASR dataset include speaker ID, age bracket, gender, language, and recording environment (e.g., Indoor, Outdoor, Office).

\vspace{-0.3cm}
\subsection{TTS Data Collection}
\label{subsec:tts_data_coll}

The TTS dataset was designed for building high-quality synthetic voices across multiple distinct studio voice actors per language. The process involved: \textbf{Script Development} of phonetically balanced scripts for each of the 15 target language configurations; \textbf{Voice Actor Selection} of 103 community participants (4--8 speakers per language: 51 male and 52 female); and \textbf{Studio Recording} in a professional environment yielding $\sim$247 hours (25,536 utterances; $\sim$16.5 hours on average per language) of clean studio audio.
The collected attributes for the TTS dataset include speaker ID, gender, and language.

\vspace{-0.2cm}
\subsection{Dataset Statistics}
\label{sec:dataset_statistics}
The WAXAL collection comprises distinct ASR and TTS datasets. The ASR dataset spans 19 languages, and the TTS dataset spans 15 language configurations (27 Sub-Saharan African languages in total across both modalities). The total size of the released ASR data is 1.7 TB, while the TTS data is 99 GB. Figure~\ref{fig:asr_trans_tts_utts_dist}, Figure~\ref{fig:asr_demo_tts_dist}, and Table~\ref{tab:waxal_overview} provide an overview; here, an instance refers to a single recorded utterance.
\begin{table}[h]
\vspace{-0.2cm}
\centering
\caption{Statistics of the WAXAL TTS and ASR datasets.}
\label{tab:waxal_overview}
\begin{tabular}{lcccc}
\toprule
\textbf{Dataset} & \textbf{\# Languages} & \textbf{\# Speakers} & \textbf{Total Hours} & \textbf{\# Utterances} \\
\midrule
WAXAL-TTS & 15 & 103 & $\sim$247  & 25,536  \\
WAXAL-ASR (Transcribed) & 19 & 5,021 & $\sim$2,340 & 438,229 \\
WAXAL-ASR (All) & 19 & 5,021+ & 14,140+ & $\sim$1,956,877 \\
\bottomrule
\end{tabular}
\vspace{-0.4cm}
\end{table}

\section{Experiments}
\label{sec:experiments}

While WAXAL encompasses both speech modalities (with TTS acoustic consistency and synthesis ablations evaluated in Appendix~\ref{app:tts_evaluation}), our primary empirical evaluation in this section focuses on the 19-language ASR corpus due to its diverse scale. To demonstrate its utility as a training and evaluation resource, we benchmark three popular ASR models representing distinct architectural paradigms.
We evaluate both zero-shot and fine-tuned performance across all 19 languages in the WAXAL-ASR dataset using strictly speaker-disjoint splits and multi-seed confidence intervals, and test whether models fine-tuned solely on WAXAL generalize to three independently collected out-of-distribution benchmarks: FLEURS, Mozilla Common Voice, and Sunbird SALT.

\vspace{-0.2cm}
\subsection{Experimental Setup}
\vspace{-0.2cm}
\label{subsec:exp_setup}

\noindent\textbf{Models.}
We select three multilingual ASR models covering distinct modeling paradigms to demonstrate that the benefits of WAXAL are universal across architectural approaches.
\textbf{Whisper-large-v3}~\cite{radford2023whisper} (\texttt{openai/whisper-large-v3}, $\sim$1.5B parameters) is an encoder-decoder model trained with large-scale supervised multilingual speech data, representing the supervised pretraining paradigm.
\textbf{MMS-1B}~\cite{pratap2024scaling} (\texttt{facebook/mms-1b-all}, $\sim$1B parameters) is built on self-supervised wav2vec~2.0 representations with language-specific CTC adapter fine-tuning, designed explicitly for low-resource and underrepresented languages.
\textbf{Gemma 3n 2B}~\cite{gemma3n2025} (\texttt{google/gemma-3n-E2B-it}, $\sim$2B effective parameters) is a multimodal large language model with a native audio encoder that encodes audio at 6.25 tokens per second; we evaluate it in a causal language modeling framework for ASR, representing the emerging paradigm of multimodal LLMs applied to speech recognition. All experiments were conducted on NVIDIA H100 GPUs.

\noindent\textbf{Speaker-Disjoint Data Splits and OOD Benchmarks.}
To ensure that ASR performance reflects true acoustic and linguistic generalization rather than speaker memorization, all 19 languages are evaluated on strictly speaker-disjoint ($\approx$80/10/10) train/validation/test splits with $0\%$ speaker overlap between training and test sets. For cross-dataset out-of-distribution (OOD) transfer evaluation, models fine-tuned exclusively on WAXAL are evaluated zero-shot on three external benchmarks: (1) \textbf{FLEURS}~\cite{conneau2023fleurs} across 6 overlapping languages (\texttt{amh}, \texttt{ful}, \texttt{lin}, \texttt{lug}, \texttt{orm}, \texttt{sna}); (2) \textbf{Mozilla Common Voice}~\cite{ardila2019common} across 4 overlapping languages (\texttt{amh}, \texttt{dag}, \texttt{lug}, \texttt{orm}); and (3) \textbf{Sunbird SALT}~\cite{akera2022salt} across 3 overlapping East African languages (\texttt{ach}, \texttt{lug}, \texttt{nyn}).

\noindent\textbf{Metrics and Text Normalization.}
We report Word Error Rate (WER) and Character Error Rate (CER) per language and macro-averaged across all evaluated languages. Prior to scoring, reference and predicted transcriptions are normalized as detailed in Appendix~\ref{app:text_norm}. Reporting CER alongside WER is critical for agglutinative African languages where single character substitutions disproportionately penalize word-level accuracy.

\noindent\textbf{Zero-Shot Protocol.}
For zero-shot evaluation, each model is assessed without any fine-tuning on the speaker-disjoint test split of each language in the WAXAL dataset. Whisper-large-v3 relies on its built-in automatic language detection. MMS-1B loads the pretrained language-specific adapter for each language as provided in the original MMS release~\cite{pratap2024scaling}; however, no pretrained adapters exist for Dagbani (\texttt{dag}) and Ikposo (\texttt{kpo}), so these two languages are excluded from MMS-1B zero-shot evaluation. Gemma 3n 2B uses a language-aware system instruction: \textit{``You are an expert model on transcribing \{language name\}.''}

\noindent\textbf{Whisper-large-v3 Fine-Tuning.}
For fine-tuning the Whisper-large-v3 model, we followed the training protocol inspired by Akera and Nafula~\cite{akera2025speechdatanecessaryasr}, who evaluated data scaling---the systematic analysis of how varying hours of training data impact transcription accuracy---for African languages.
We performed full fine-tuning of all model parameters, without LoRA. The model was trained for a maximum of $20,000$ steps with a learning rate of $10^{-5}$, a maximum sequence length of $448$, a per-device batch size of $4$, and $8$ gradient accumulation steps. We used early stopping with a patience of $4$ evaluations.

\noindent\textbf{MMS-1B Fine-Tuning.}
For the MMS-1B model, we followed the adapter-based fine-tuning protocol described by Hugging Face for MMS adapters~\cite{mms_adapters_blog}. We re-initialized the language adapters to align the output vocabulary with each language's character set in WAXAL, preventing out-of-vocabulary errors during transcription. The model was trained for a maximum of $1,500$ steps with a learning rate of $10^{-3}$, a warmup of $100$ steps, a per-device batch size of $4$, and $8$ gradient accumulation steps.

\noindent\textbf{Gemma 3n 2B Fine-Tuning.}
For the Gemma 3n 2B model, we adopted a causal language modeling approach for transcription. The model was prompted with a system message: \textit{``You are an assistant that transcribes speech accurately."} and a user prompt: \textit{``Please transcribe this audio."}
Due to the model size, we employed Low-Rank Adaptation (LoRA) targeting the attention projection layers ($q, k, v, o$), feed-forward modules ($gate, up, down$), and the multi-modal projector. We used a high LoRA rank of $r = 1024$ and $\alpha = 2048$ with no dropout, as preliminary experiments showed that lower ranks yielded insufficient model capacity for the transcription task.
The model was trained for a maximum of $5,000$ steps with a learning rate of $10^{-4}$, a per-device batch size of $4$, and $4$ gradient accumulation steps.

\noindent\textbf{Statistical Rigor and Multi-Seed Evaluation.}
To quantify training variance across all 57 model--language combinations ($3\text{ models} \times 19\text{ languages}$), we perform $N \ge 3$ independent fine-tuning runs per language ($N \in [3, 9]$, totaling $192$ fine-tuned models; exact run counts per language are listed in Table~\ref{tab:waxal_cer_appendix}) and report $\text{Mean} \pm \text{95\% CI}$ computed via the Student-$t$ distribution ($t_{0.975, N-1} \cdot s / \sqrt{N}$). In addition to cross-run confidence intervals, we conduct paired utterance-level bootstrap resampling ($10{,}000$ iterations per language) comparing zero-shot and fine-tuned utterance error rates to verify statistical significance ($p < 0.0001$).

\begin{figure}[!thbp]
    \centering
    \vspace{-0.2cm}
    \includegraphics[width=\linewidth]{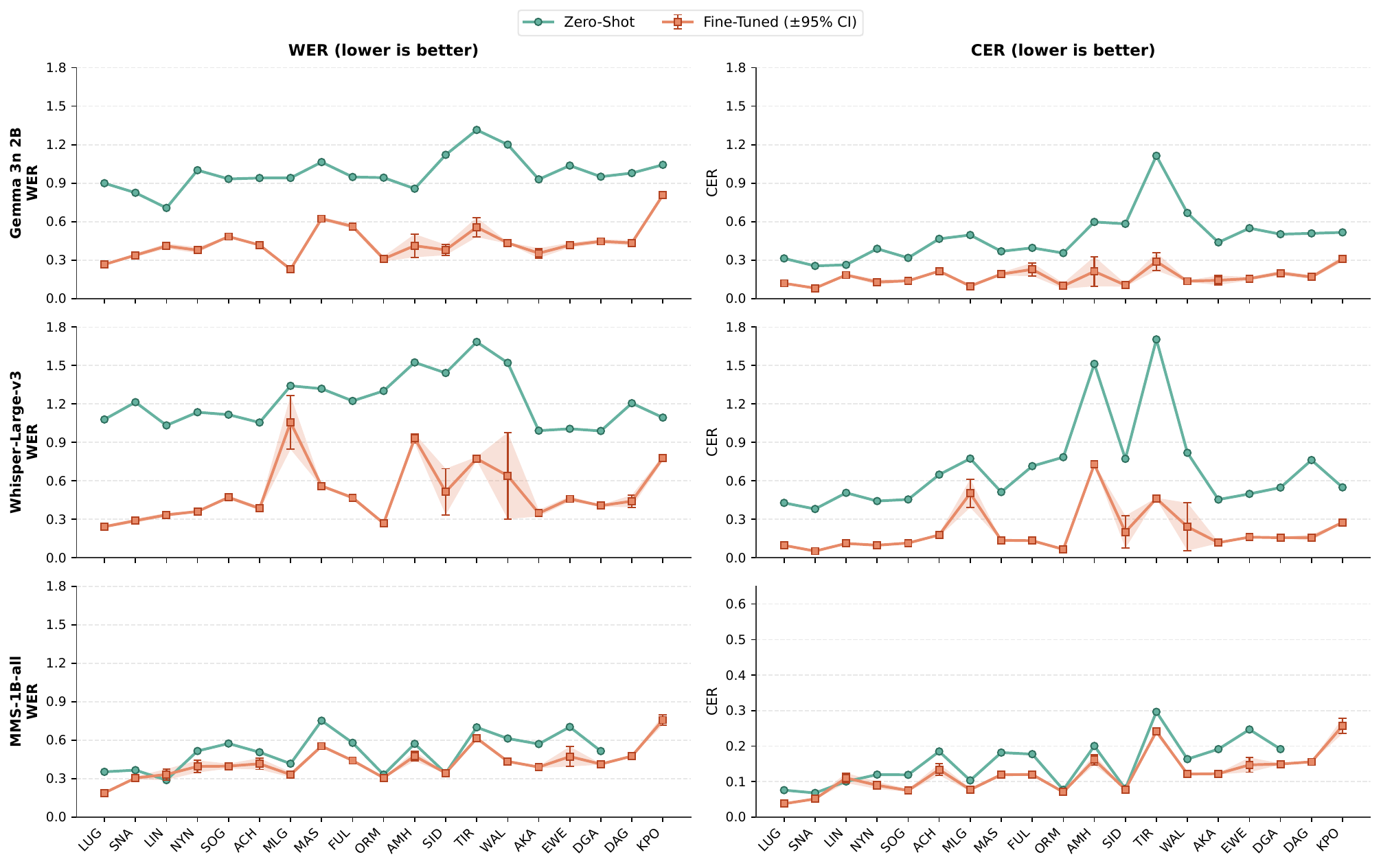}
     \caption{Zero-shot vs.\ fine-tuned Word Error Rate (WER) and Character Error Rate (CER) per language on the speaker-disjoint WAXAL test sets (\texttt{v2}) across three models. %
    Teal lines (\textcolor[HTML]{3B9683}{$\bullet$}) show Zero-Shot error rates; coral lines (\textcolor[HTML]{C95533}{\rule{0.55em}{0.55em}}) show Fine-Tuned error rates ($\text{Mean} \pm 95\%$ CI over $N \ge 3$ runs per language). %
    Fine-tuning consistently and significantly reduces both WER and CER across models and languages. Full numerical tables with 95\% CIs are provided in Table~\ref{tab:asr_results} (WER) and Table~\ref{tab:waxal_cer_appendix} (CER) in Appendix~\ref{app:supp_asr}.}
    \label{fig:waxal_zs_ft_per_lang}
    \vspace{-0.5cm}
\end{figure}

\subsection{Results on WAXAL}

Figure~\ref{fig:waxal_zs_ft_per_lang} presents the zero-shot and fine-tuned WER and CER (with $\pm 95\%$ confidence intervals) for all three models across the 19 WAXAL languages on the strictly speaker-disjoint test splits. Fine-tuning substantially reduces error rates across all models and languages, with full per-language WER and CER tables provided in Table~\ref{tab:asr_results} and Table~\ref{tab:waxal_cer_appendix} in Appendix~\ref{app:supp_asr}.

\noindent\textbf{Macro-Average WER and CER Improvements.}
Aggregated across all 19 speaker-disjoint test sets ($192$ fine-tuned runs total), monolingual fine-tuning on WAXAL reduces macro-average WER from $0.9806$ to $\mathbf{0.4342 \pm 0.0231}$ ($-55.7\%$ relative reduction) and CER from $0.4783$ to $\mathbf{0.1678 \pm 0.0240}$ ($-64.9\%$ relative) for \textbf{Gemma 3n 2B}; reduces WER from $1.2250$ to $\mathbf{0.5122 \pm 0.0530}$ ($-58.2\%$ relative) and CER from $0.6983$ to $\mathbf{0.2104 \pm 0.0268}$ ($-69.9\%$ relative) for \textbf{Whisper-Large-v3}; and reduces WER from $0.5115$ (across the 17 languages with pretrained zero-shot adapters) to $\mathbf{0.4233 \pm 0.0213}$ ($-17.2\%$ relative across all 19 languages) and CER from $0.1515$ to $\mathbf{0.1219 \pm 0.0068}$ ($-19.5\%$ relative) for \textbf{MMS-1B-all}. Moreover, fine-tuning \texttt{MMS-1B-all} on WAXAL unlocks transcription for Dagbani (\texttt{dag}: $0.4754 \pm 0.0027$ WER, $0.1555 \pm 0.0008$ CER) and Ikposo (\texttt{kpo}: $0.7574 \pm 0.0384$ WER, $0.2570 \pm 0.0219$ CER), neither of which is supported in the official zero-shot MMS release. Across all 57 model--language pairs, paired utterance-level bootstrap resampling ($10{,}000$ iterations) confirms that fine-tuning improvements over zero-shot baselines are statistically significant at $p < 0.0001$, with tight run-to-run variance ($\pm 0.0015$ to $\pm 0.0475$ WER CI on most languages).

\vspace{-0.2cm}
\subsection{Out-of-Distribution Generalization}
\vspace{-0.2cm}
\label{sec:fleurs}

\begin{figure}[bhtp]
\vspace{-0.2cm}
    \centering
    \includegraphics[width=\linewidth]{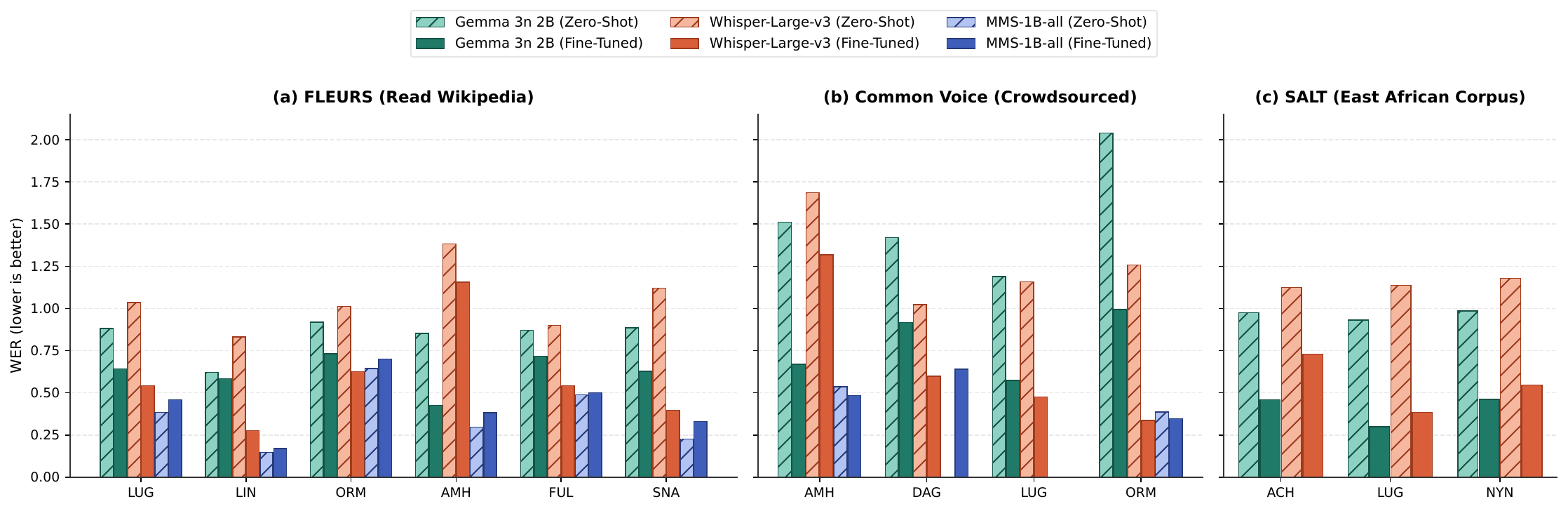}
    \caption{\textbf{Out-of-Distribution (OOD) Generalization across Three External Benchmarks:} \textbf{FLEURS v2} (6 languages), \textbf{Mozilla Common Voice} (4 languages), and \textbf{Sunbird SALT} (3 East African languages). Models are fine-tuned solely on speaker-disjoint WAXAL (\texttt{v2}) splits without exposure to target benchmark data. Note that FLEURS is in-domain for zero-shot \texttt{MMS-1B-all}, whereas on truly external benchmarks (\textbf{Common Voice}), WAXAL fine-tuning reduces WER across all three models including \texttt{MMS-1B-all}. Detailed WER and CER are provided in Table~\ref{tab:ood_all_appendix} in Appendix~\ref{app:supp_asr}.}
    \label{fig:fleurs_genz_hr}
    \vspace{-0.3cm}
\end{figure}

A critical test for any speech corpus is whether models trained on it generalize zero-shot to independently collected benchmarks with unseen speakers, recording equipment, and elicitation domains.
We evaluate our speaker-disjoint WAXAL-fine-tuned checkpoints across three external benchmarks: \textbf{FLEURS}~\cite{conneau2023fleurs}, \textbf{Mozilla Common Voice}~\cite{ardila2019common}, and \textbf{Sunbird SALT}~\cite{akera2022salt}. Figure~\ref{fig:fleurs_genz_hr} summarizes the OOD results across all three benchmarks, with complete WER and CER numbers reported in Table~\ref{tab:ood_all_appendix} in Appendix~\ref{app:supp_asr}.

\noindent\textbf{Generalization to FLEURS (6 Languages).}
Across the six overlapping languages in FLEURS (\texttt{lug}, \texttt{lin}, \texttt{orm}, \texttt{amh}, \texttt{ful}, \texttt{sna}), where audio consists of scripted Wikipedia readings rather than spontaneous image descriptions, fine-tuning solely on WAXAL reduces macro-average WER from $0.8383$ to $\mathbf{0.6211}$ (CER from $0.3560$ to $\mathbf{0.2667}$) for Gemma 3n 2B and from $1.0469$ to $\mathbf{0.5900}$ (CER from $0.5430$ to $\mathbf{0.2980}$) for Whisper-Large-v3, improving performance on every single language (e.g., Whisper WER drops from $0.8306 \rightarrow \mathbf{0.2771}$ on Lingala and $1.1200 \rightarrow \mathbf{0.3960}$ on Shona; Gemma 3n drops from $0.8515 \rightarrow \mathbf{0.4252}$ on Amharic). For \texttt{MMS-1B-all}, we note that Meta's released checkpoint was trained directly on the FLEURS training splits~\cite{pratap2024scaling}, making FLEURS \textit{in-domain} for zero-shot \texttt{MMS-1B-all} ($0.3646$ WER); re-initializing its adapter layers for WAXAL fine-tuning replaces those FLEURS-specific adapter weights yet still maintains competitive OOD accuracy ($0.4239$ WER, $0.1124$ CER).

\noindent\textbf{Generalization to Mozilla Common Voice (4 Languages) and Sunbird SALT (3 Languages).}
When evaluated on external benchmarks that were \textit{not} included in MMS adapter pretraining, WAXAL fine-tuning improves \textbf{all three models}, including \texttt{MMS-1B-all}. On \textbf{Mozilla Common Voice} (\texttt{amh}, \texttt{dag}, \texttt{lug}, \texttt{orm}), WAXAL fine-tuning reduces macro-average WER from $1.5404$ to $\mathbf{0.7881}$ ($-48.8\%$ relative; CER $0.9661 \rightarrow \mathbf{0.4467}$) for Gemma 3n 2B and from $1.2809$ to $\mathbf{0.6823}$ ($-46.7\%$ relative; CER $0.9592 \rightarrow \mathbf{0.3862}$) for Whisper-Large-v3 (with Oromo dropping from $1.2567 \rightarrow \mathbf{0.3354}$ WER). Crucially, \texttt{MMS-1B-all} also improves across Common Voice: WER drops from $0.5351 \rightarrow \mathbf{0.4846}$ on Amharic and from $0.3859 \rightarrow \mathbf{0.3472}$ on Oromo, while WAXAL fine-tuning unlocks Dagbani on Common Voice ($\mathbf{0.6404}$ WER, $0.1672$ CER). Similarly, on the \textbf{Sunbird SALT} benchmark (\texttt{ach}, \texttt{lug}, \texttt{nyn}), WAXAL fine-tuning reduces macro-average WER from $0.9639$ to $\mathbf{0.4064}$ ($-57.8\%$ relative; Luganda $0.9312 \rightarrow \mathbf{0.2987}$) for Gemma 3n 2B and from $1.1464$ to $\mathbf{0.5530}$ ($-51.8\%$ relative; Luganda $1.1374 \rightarrow \mathbf{0.3830}$) for Whisper-Large-v3.

\noindent\textbf{Bidirectional Cross-Dataset Transfer (\textsc{WAXAL} $\leftrightarrow$ FLEURS).}
Comparing \texttt{MMS-1B} adapters fine-tuned on WAXAL versus FLEURS across the six overlapping languages (Table~\ref{tab:cross_train} in Appendix~\ref{app:supp_asr}), the mean in-domain vs.\ cross-domain WER gap is modestly larger on spontaneous WAXAL speech ($-0.07$, up to $-0.16$ on Luganda and $-0.14$ on Fula) than on scripted FLEURS speech ($-0.06$).

\vspace{-0.1cm}
\subsection{LLM-as-Judge: Meaning Preservation}
\label{sec:llm_judge}

WER and CER measure surface-level string differences but do not capture whether a transcript preserves the intended meaning for a native listener.
To complement these metrics, we conduct a model-based semantic evaluation of the Gemma 3n 2B fine-tuned predictions across all 19 languages using Gemini 3.1 Pro as the judge model, and subsequently validate the judge's scores against native human annotations.

\noindent\textbf{Setup.} The judge is given the target language, the reference transcription, and the ASR prediction, and is asked to score meaning preservation.
We evaluate two scoring rubrics (see Appendix~\ref{app:llm_judge_prompts} for the full prompts):
\begin{itemize}[noitemsep,nolistsep,leftmargin=8pt]
    \item \textbf{Discrete scoring:} The judge assigns one of three scores---$1.0$ (meaning fully preserved), $0.5$ (partial preservation), or $0.0$ (meaning lost)---following an expert-linguist rubric that distinguishes benign dialectal/orthographic variation from critical hallucinations.
    \item \textbf{Continuous scoring:} The judge assigns a float in $[0.0, 1.0]$ using an Information Transfer Rate framework that differentiates primary intent from secondary details and penalizes hallucinations more severely than omissions.
\end{itemize}
Each language is evaluated on 50\% of its test examples (19,903 total). The evaluation is run independently for each scoring mode.

\begin{figure}[!thbp]
\vspace{-0.3cm}
  \centering
  \includegraphics[width=\linewidth]{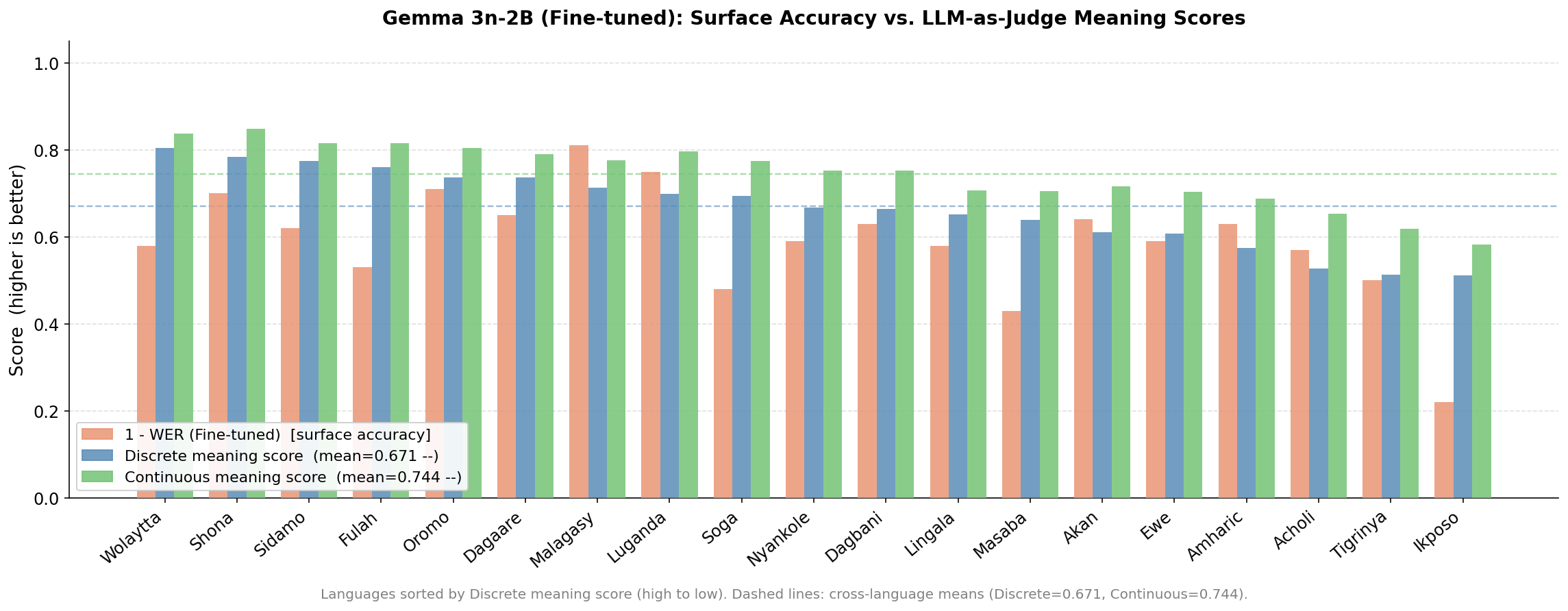}
  \caption{Gemma 3n 2B (fine-tuned): surface accuracy ($1 - \text{WER}$) vs.\ LLM-as-judge meaning-preservation scores per language. Languages are sorted by Discrete score (high to low). Dashed lines mark the cross-language means: Discrete~$= 0.667$ and Continuous~$= 0.744$. Higher is better for all three metrics. Continuous scores consistently exceed Discrete scores ($+0.077$ on average), reflecting partial-credit awards that Discrete scoring quantizes away. Languages with lower WER (e.g., Wolaytta, Shona) also rank highest on both semantic rubrics. Detailed Discrete ($\in\{0.0, 0.5, 1.0\}$) and Continuous ($\in[0.0, 1.0]$) scores are shown in Table~\ref{tab:llm_judge} in Appendix~\ref{sec:app_llm_judge}.}
  \label{fig:llm_judge}
  \vspace{-0.4cm}
\end{figure}

\noindent\textbf{Results.} Figure~\ref{fig:llm_judge} illustrates the per-language scores, with precise details given in Table~\ref{tab:llm_judge} in Appendix~\ref{sec:app_llm_judge}.
Continuous scores are consistently higher than discrete scores (macro-average $+0.077$), as expected: the continuous rubric awards partial credit that the discrete rubric quantizes into the lower bucket.
Overall, the fine-tuned Gemma 3n 2B model achieves a macro-average discrete score of $0.667$ and a macro-average continuous score of $0.744$ across all 19 languages, indicating that the majority of meaning is preserved in the predicted transcripts despite imperfect surface-level accuracy.
Languages with lower WER (e.g., Wolaytta, Shona) also score highest on both meaning-preservation rubrics, while the most challenging languages (Acholi, Amharic, Tigrinya) show the largest deltas between discrete and continuous scoring, reflecting many partially correct transcriptions that WER does not capture.

\noindent\textbf{Human Validation.} To verify that Gemini 3.1 Pro provides reliable semantic judgments in low-resource African languages, we conducted a meta-evaluation with native speakers of Akan and Amharic. For each language, we randomly sampled 50 examples from the evaluation set ($N = 99$ valid reference--prediction pairs in total: $50$ for Akan and $49$ for Amharic) and asked native speakers to independently score each pair on the same continuous $[0.0, 1.0]$ scale used by the LLM judge. Comparing human and LLM scores using Pearson correlation ($r$), Mean Absolute Error (MAE), and mean bias ($\text{LLM} - \text{Human}$), the LLM judge achieves strong agreement across both languages: $r = 0.74$ ($\text{MAE} = 0.13$, mean bias $= +0.01$) for Akan and $r = 0.68$ ($\text{MAE} = 0.14$, mean bias $= +0.05$) for Amharic, with an overall correlation of $r = 0.71$, $\text{MAE} = 0.14$, and mean bias $= +0.03$ across all $N = 99$ samples (see Table~\ref{tab:human_llm_agreement} in Appendix~\ref{sec:app_llm_judge}). These results confirm that the LLM judge serves as a reliable proxy for native human evaluation with minimal systematic tendency to inflate or deflate scores.

\vspace{-0.4cm}
\section{Discussions, Future Directions and Limitations}

\vspace{-0.2cm}
\textbf{Whisper-Large-v3 Failure Modes.}
\label{subsec:whisper_failures}
While fine-tuning consistently improves WER across all models and languages, our analysis reveals that Whisper-Large-v3 exhibits qualitatively distinct failure modes on Amharic ($0.9314 \pm 0.0365$ WER) and Malagasy ($1.0560 \pm 0.2094$ WER) that aggregate WER does not fully capture. For \textbf{Amharic}, the zero-shot model fails primarily through wrong-script outputs (50.8\% of utterances) and hallucinations (49.2\%; Table~\ref{tab:whisper_amh_zeroshot}), indicating that Amharic's Ethiopic (\textit{Ge'ez}) script is largely absent from the model's pretraining data. Fine-tuning eliminates wrong-script outputs and nearly halves hallucinations (49.2\% $\to$ 25.8\%), but introduces a new dominant failure: Unicode replacement characters (\texttt{U+FFFD}) appear in 55.6\% of fine-tuned predictions (Table~\ref{tab:whisper_amh_finetuned}) because Whisper's byte-level BPE tokenizer splits 3-byte UTF-8 Ethiopic characters across subword boundaries, and words are frequently truncated mid-syllable (e.g., \textit{``\textethi{መድኃኒት}''} [medhanit] $\to$ \textit{``\textethi{መድ}''} [med]). For \textbf{Malagasy}, fine-tuning swaps one catastrophic failure mode for another: 76.8\% hallucination (wrong language or gibberish) in zero-shot becomes $\sim$87\% autoregressive repetition or stuttering (\texttt{ny ny ny\ldots}) in the fine-tuned model, with only 0.6\% of samples genuinely corrected. In contrast, character-CTC (\texttt{MMS-1B-all}: $0.4761$ Amharic, $0.3314$ Malagasy WER) and native multimodal LLM (\texttt{Gemma 3n 2B}: $0.4120$ Amharic, $0.2281$ Malagasy WER) architectures avoid both byte-splitting corruption and autoregressive stuttering loops. Additional details and examples are provided in Appendix~\ref{app:whisper_failures}.

\textbf{Comments on WAXAL-TTS.} While the main-text experiments focus on the 19-language ASR corpus, WAXAL also provides a rich studio-quality TTS dataset across 15 language configurations ($\sim$247 hours, 25,536 utterances, 103 speakers). Because neural TTS synthesis architectures require utterances to be segmented and force-aligned into shorter durations ($\le 10$--$15$\,s) with strict text--audio alignment, we provide a dedicated evaluation of WAXAL-TTS in Appendix~\ref{app:tts_evaluation}, including an ECAPA-TDNN speaker-consistency and voice-diversity audit across the 15 TTS configurations ($0.793$ mean intra-speaker cosine similarity across the 13 evaluated corpora and $0.25$ mean median pairwise cosine similarity across speakers) and a controlled 4-arm \texttt{MMS-TTS} training ablation on Acholi, Kiswahili, and Luganda showing that filtering studio segments at $\ge 95\%$ forced-alignment confidence reduces round-trip synthesis CER by up to $29\%$ relative to baseline \texttt{MMS-TTS}.

\textbf{Qualitative Examples: LLM-as-Judge Scoring.}

Table~\ref{tab:qual_examples_akan4} presents representative Akan examples from the LLM-as-judge evaluation of Gemma 3n 2B predictions, illustrating the three scoring tiers, the subtlety of the judge, and the corresponding failure modes. Extended examples for both Akan and Amharic---two languages with large Discrete--Continuous score gaps ($+0.105$ and $+0.114$, respectively)---are provided in Appendix~\ref{sec:app_llm_judge} (Table~\ref{tab:qual_examples}). The qualitative analysis reveals that the zero-shot model frequently suffers from severe cross-script hallucinations, inappropriately injecting characters from other scripts. While fine-tuning effectively eliminates these, the judge's partial scores (0.5) illustrate how the adapted model still occasionally struggles with minor lexical substitutions, such as altering pronouns or shifting secondary spatial details.

\afterpage{
\clearpage
\begingroup
\scriptsize
\setlength{\tabcolsep}{2pt}
\setlength{\LTcapwidth}{\textwidth}
\begin{xltabular}{\textwidth}{c X X X}
\caption{Qualitative examples of Gemma 3n 2B zero-shot and fine-tuned predictions for Akan, with WER, Discrete (D), and Continuous (C) scores from the LLM-as-judge.}\label{tab:qual_examples_akan4}\\
\toprule
\textbf{Score} & \textbf{Reference} & \textbf{Zero shot prediction (WER)} & \textbf{Fine tuned prediction (WER)} \\
\midrule
\endfirsthead
\caption[]{Qualitative examples (continued)}\\
\toprule
\textbf{Score} & \textbf{Reference} & \textbf{Zero shot prediction (WER)} & \textbf{Fine tuned prediction (WER)} \\
\midrule
\endhead
\multicolumn{4}{r}{Continued on next page} \\
\bottomrule
\endfoot
\bottomrule
\endlastfoot
\midrule
\makecell[t]{D:0.0\\C:0.1} & Mmaa sɛ wogyina gyina hɔ. Ebi de ne nsa asɔ ne sisi. Ebi ɛde wɔn nsa aka abom. Ebi egyina hɔ ɛde wɔn nsa egu wɔn bo. Ebi nso ɛsere. Ataade a ɛhyɛ ebinom ɛyɛ fitaa. ...%
& Enma, bi se fo kenee gya ho. Ebi ti nsa aso wo nsisi. Ebi atie wo nsa akan wo. Ebi nso esere atalia ehu ebini ye fita. Ebi de ahamono ne bi denso yekoko. (WER: 85\%) & Mmaa bi hyɛ adwadipono ho. Ebi de ne nsa asɔ wɔn sisi. Ebi de wɔn nsa aka aboom. Ebi akyi ahokɔsradeɛ nso wɔ wɔn ho. (WER: 71\%) \\*
\multicolumn{4}{p{\dimexpr\textwidth-2\tabcolsep\relax}}{\textbf{Explanation:} The reference describes women standing ('Mmaa sɛ wogyina gyina hɔ')... The prediction fundamentally changes the meaning right from the start. It says 'Mmaa bi hyɛ adwadipono ho' (some women are at a shop/counter), replacing the description of them standing.} \\
\midrule
\makecell[t]{D:0.5\\C:0.85} & Mmayewa mmienu, wɔn mmienu nyinaa hyɛ kyɛ. Kanea asɔ wɔ faako a wɔgyina no, na atosodeɛ sisi baabi a wɔgyina no. Deɛ ɔgyina anim no kuta bɔtɔ. & Ɛ\textethi{ማ }wo ameyinu, omue nyinaa a hyɛkye, kannia a sɔwɔ faa akun wo gyina no. Na atosodie esisi baabi a wo mu gyina no, diee ogyinaa nimu no ekuta bot\texttibetan{◌ོ}. (WER: 85\%) & Mmayewa mmienu wɔ mmienu nyinaa hyɛ kyɛ. Kanea asɔ wɔ faako a ogyina no, na atosodeɛ sisi baabi a wɔgyina no. Deɛ ɔgyina anim no kuta bɔtɔ. (WER: 7\%) \\*
\multicolumn{4}{p{\dimexpr\textwidth-2\tabcolsep\relax}}{\textbf{Explanation:} The ZS prediction contains severe phonetic errors and cross-script hallucinations (including Ethiopic and Tibetan). The FT prediction successfully reconstructs the vocabulary to match the reference but incorrectly substitutes the plural pronoun subject prefix in ``wɔgyina'' with the singular ``o'' in ``ogyina'', leading to a partial score.} \\
\midrule
\makecell[t]{D:0.5\\C:0.5} & Adɔfofoɔ ne amanfoɔ a wɔahyia mu a wɔrehwɛ sini. Abusua baako a wɔyɛ nnan rehwɛ wɔn sini, ɛna amanfoɔ pii nso so a wɔahyia mu wɔ sinitwabea, wɔn nso so rehwɛ wɔn sini. & A dafo fo, eny aman fo a, won ehyie mu a won hweene. Ebusia baa h\texttibetan{◌ོ}, a won yɛ enai ehɛ won sinie. En aman fo pi isu so a ehɛ mu ewo sinie trabia wonzo se ehɛ won sina? (WER: 103\%) & Adɔfofoɔ ne amanfoɔ a wɔahyia mu a wɔrehwɛ sini. Abusua baako a wɔyɛ nnan rehyɛ wɔn sini. Ɛnna amanfoɔ pii nso so ahyia mu wɔ sini. N'atwa beaeɛ a wɔn nso so rehyɛ wɔn sini. (WER: 34\%) \\*
\multicolumn{4}{p{\dimexpr\textwidth-2\tabcolsep\relax}}{\textbf{Explanation:} The ZS prediction is heavily fragmented with critical phonetic errors and Tibetan script hallucinations. The FT model successfully reconstructs the sentence structure and core meaning, though it introduces minor lexical substitutions (e.g., ``rehyɛ'' instead of ``rehwɛ'') that fundamentally alter the actions and locations affecting the second and third sentences.} \\
\midrule
\makecell[t]{D:1.0\\C:1.0} & Nnipa bebree wɔ ha. Nipa no bi rebɔ nhwene. Nipa no bi hyɛ fugu. Nipa no bi resa. Nnipa no bi nso gyina hɔ rehwɛ. Nnua bebree gu wɔn akyi. & Epa bebrewa. Nipa ne bi ɛbɔnshwini. Nipa ne bi shɛfugu. Nipa ne bi ɛsɛ. Nipa ne bi ensu gyina oɔ ɛhɛ. Enia bebreɛgu wo omaiki. (WER: 70\%) & Nnipa bebree wɔ ha. Nnipa no bi rebɔ nhwene. Nnipa no bi hyɛ fugu. Nnipa no bi resa. Nnipa no bi nso gyina hɔ rehwɛ. Nnua bebree gu wɔn akyi. (WER: 37\%) \\*
\multicolumn{4}{p{\dimexpr\textwidth-2\tabcolsep\relax}}{\textbf{Explanation:} The ZS model struggles heavily with poor phonetic spelling and word fragmentation. The FT model resolves these entirely, generating a nearly flawless transcription where the only difference is the benign pluralization of "Nipa" to "Nnipa".} \\
\end{xltabular}
\vspace{-0.2cm}
\endgroup
}

\vspace{-0.2cm}
\subsection{Limitations and Considerations}
\label{subsec:limitations}
The release of any large-scale human data requires careful consideration of its limitations. \textbf{Transcription Coverage} is currently limited to approximately 10--17\% of the total audio collected. \textbf{Dialectal Representation} may not capture full variation despite efforts to recruit diverse speakers. There is a small risk of \textbf{Unintended Content} in unscripted data, mitigated by manual quality control. Finally, \textbf{Use-Case Specificity} means the ASR dataset is not suited for training high-quality single-speaker TTS models.

\vspace{-0.1cm}
\subsection{Ethical Considerations}
\label{subsec:ethical_considerations}
\vspace{-0.1cm}
Regarding ethical considerations, \textbf{Informed Consent} was obtained from all participants and voice actors for research release under a permissive license. \textbf{Speaker Privacy} was protected by removing PII, though accent and language might still imply background attributes. \textbf{Voice Actor Rights} are balanced against broad community benefit, and we believe enabling technology outweighs unforeseen risks. Finally, \textbf{Annotator Compensation} for local experts was set well above the local hourly average.

\vspace{-0.2cm}
\section{Conclusion}
\vspace{-0.2cm}
We have introduced WAXAL, a large-scale speech corpus comprising approximately 2,340 hours of transcribed ASR data from 5,021 speakers (over 14,100 hours of total collected audio) across 19 languages and $\sim$247 hours of studio-quality TTS recordings (25,536 utterances from 103 speakers) across 15 languages, spanning 27 Sub-Saharan African languages in total. Crucially, this project was executed through a \textbf{community-owned partnership model} with five African academic and community organizations. This approach not only prioritized local expertise and ethical data handling, but also actively built the local ecosystem and seeded sustainable data collection capabilities, \textbf{demonstrating a scalable framework for future low-resource language resource development}.

Our multi-seed benchmarking experiments ($192$ fine-tuned runs across 19 speaker-disjoint splits) demonstrate WAXAL's practical value as a training resource. Fine-tuning three architecturally diverse ASR models---Gemma 3n 2B, Whisper-Large-v3, and MMS-1B-all---yields consistent, statistically significant improvements, with macro-average WER reductions of $56\%$, $58\%$, and $17\%$ respectively, alongside CER reductions of up to $70\%$. 
Importantly, models fine-tuned exclusively on WAXAL generalize zero-shot to three independently collected external benchmarks: FLEURS (up to $67\%$ WER reduction on Lingala: $0.8306 \rightarrow 0.2771$), Mozilla Common Voice ($49\%$ macro-average WER reduction for Gemma 3n and $47\%$ for Whisper, while improving MMS-1B-all across all evaluated languages), and Sunbird SALT ($58\%$ macro-average WER reduction for Gemma 3n). Beyond surface-level error rates, the LLM-as-judge's mean continuous score of $0.744$ across all 19 languages indicates that much of the core semantic meaning is preserved in the fine-tuned predictions.

We hope that this resource will empower researchers and developers to build inclusive, practical speech technologies for the communities these languages represent. The WAXAL dataset is released publicly under the CC-BY-4.0 license at \href{https://huggingface.co/datasets/google/WaxalNLP}{https://huggingface.co/datasets/google/WaxalNLP}.

\vfill\pagebreak

\bibliographystyle{IEEEbib}
\bibliography{refs}

\newpage
\appendix

\section{Supplementary Dataset Details and ASR Results}
\label{app:supp_asr}

This appendix provides complete dataset statistics and full numerical results for the speaker-disjoint ASR benchmarks, multi-seed confidence intervals, speaker-leakage ablation, and out-of-distribution generalization experiments:
\begin{itemize}[noitemsep,nolistsep,leftmargin=12pt]
    \item \textbf{Section~\ref{app:text_norm}:} Complete text normalization procedures, Ge'ez script handling, and linguistic rationale for CER reporting.
    \item \textbf{Table~\ref{tab:asr_stats}:} Per-language speaker counts, utterance counts, and duration statistics for WAXAL-ASR (\texttt{v2} speaker-disjoint splits).
    \item \textbf{Table~\ref{tab:asr_results}:} Word Error Rate (WER, $\text{Mean} \pm \text{95\% CI}$ over $N \ge 3$ runs) across all 19 languages for Gemma 3n 2B, Whisper-Large-v3, and MMS-1B-all.
    \item \textbf{Table~\ref{tab:waxal_cer_appendix}:} Character Error Rate (CER, $\text{Mean} \pm \text{95\% CI}$) and exact number of independent training runs ($N$) per language across all 19 languages ($192$ runs total).
    \item \textbf{Section~\ref{app:leakage_ablation} \& Table~\ref{tab:leakage_ablation}:} Speaker-leakage ablation comparing fine-tuned WER on the initial utterance-random splits (\texttt{v1}) versus strictly speaker-disjoint splits (\texttt{v2}) across the 13 re-split Makerere and Clear Global languages.
    \item \textbf{Table~\ref{tab:ood_all_appendix}:} Out-of-distribution (OOD) WER and CER across \textbf{FLEURS v2} (6 languages), \textbf{Mozilla Common Voice} (\texttt{v1 + lg}, 4 languages), and \textbf{Sunbird SALT} (3 languages).
    \item \textbf{Table~\ref{tab:cross_train}:} Bidirectional cross-dataset evaluation and generalization gap (\textsc{WAXAL} $\leftrightarrow$ FLEURS) with \texttt{MMS-1B} adapters across the 6 overlapping languages.
\end{itemize}

\subsection{Text Normalization and Linguistic Evaluation Protocol}
\label{app:text_norm}

Prior to scoring transcription accuracy, both reference ground-truth and model-predicted transcripts are subjected to standardized text normalization. Reference and predicted texts are converted to lowercase, stripped of standard ASCII punctuation characters (\texttt{string.punctuation}), and tokenized on whitespace.

\noindent\textbf{Ge'ez Script Normalization.}
For the two Ge'ez-script (\textit{abugida}) languages---Amharic (\texttt{amh}) and Tigrinya (\texttt{tir})---the traditional Ethiopic two-dot word separator \textit{hulet neteb} (\textethi{፡}, \texttt{U+1361}) is normalized to a standard whitespace character (` ') prior to punctuation removal. In contemporary printed and digital Ethiopic orthography, \textethi{፡} and standard blank spaces are used interchangeably as word boundaries. Treating \textethi{፡} simply as punctuation to be stripped without whitespace replacement would artificially concatenate adjacent words into single invalid compound tokens, drastically inflating Word Error Rate. Replacing \textethi{፡} with whitespace ensures robust, linguistically accurate word tokenization.

\noindent\textbf{Linguistic Rationale for CER Reporting.}
Reporting Character Error Rate (CER) alongside Word Error Rate (WER) is essential for evaluating speech recognition in Sub-Saharan African languages. Many Niger-Congo languages (specifically Bantu languages such as Luganda \texttt{lug}, Runyankole \texttt{nyn}, Soga \texttt{sog}, Masaaba \texttt{mas}, Shona \texttt{sna}, and Lingala \texttt{lin}) and Atlantic languages (e.g., Fulani \texttt{ful}) exhibit highly agglutinative and polysynthetic morphological structures. In these languages, single orthographic words routinely encode complex combinations of subject noun-class concords, tense-aspect-mood markers, object infixes, derivational suffixes, and clitics. Under standard WER evaluation, an ASR hypothesis that misrecognizes even a single phoneme or grammatical affix in an otherwise correct multi-morpheme verbal complex is penalized as an entire word failure (100\% error for that token). CER provides a critical, unbiased metric capturing fine-grained phonetic, acoustic, and sub-word morphological recognition quality that WER systematically obscures.

\begin{table}[h!]
  \centering
  \footnotesize
  \caption{Transcribed WAXAL-ASR dataset statistics per language under the strictly speaker-disjoint splits (\texttt{v1} for the 6 Voice Donor languages ($\dagger$), \texttt{v2} for the 13 re-split Makerere and Clear Global languages ($\star$), guaranteeing $0\%$ train--test speaker overlap across all 19 languages).}
  \label{tab:asr_stats}
  \begin{tabular}{l l r r r r r r}
    \toprule
    \textbf{Language} & \textbf{ISO} & \textbf{\# Speakers} & \textbf{Train} & \textbf{Val} & \textbf{Test} & \textbf{Total Utt.} & \textbf{Hours} \\
    \midrule
    Acholi$^{\star}$    & \texttt{ach} &   170 &  4,108 &   519 &   528 &  5,155 &  32.3 \\
    Oromo$^{\dagger}$   & \texttt{orm} &   441 & 38,185 & 3,078 & 3,782 & 45,045 & 220.0 \\
    Akan$^{\dagger}$    & \texttt{aka} &   404 & 10,107 & 1,123 & 1,522 & 12,752 & 102.0 \\
    Amharic$^{\dagger}$ & \texttt{amh} &   432 & 38,022 & 2,912 & 3,420 & 44,354 & 218.0 \\
    Dagaare$^{\star}$   & \texttt{dga} &   240 & 15,071 & 1,892 & 1,910 & 18,873 & 104.7 \\
    Dagbani$^{\star}$   & \texttt{dag} &   235 & 14,231 & 1,750 & 1,838 & 17,819 &  98.5 \\
    Ewe$^{\star}$       & \texttt{ewe} &   238 & 15,052 & 1,916 & 1,891 & 18,859 &  99.7 \\
    Fulani$^{\star}$    & \texttt{ful} &   215 & 19,132 & 2,358 & 2,391 & 23,881 & 124.4 \\
    Ikposo$^{\star}$    & \texttt{kpo} &   228 & 14,415 & 1,760 & 1,845 & 18,020 & 103.8 \\
    Lingala$^{\star}$   & \texttt{lin} &   205 & 14,400 & 1,844 & 1,866 & 18,110 & 101.5 \\
    Luganda$^{\star}$   & \texttt{lug} &   180 &  5,455 &   664 &   638 &  6,757 &  46.0 \\
    Malagasy$^{\star}$  & \texttt{mlg} &   210 & 18,526 & 2,270 & 2,290 & 23,086 & 182.5 \\
    Masaaba$^{\star}$   & \texttt{mas} &   165 &  6,867 &   849 &   858 &  8,574 &  48.8 \\
    Runyankole$^{\star}$& \texttt{nyn} &   172 &  6,783 &   831 &   858 &  8,472 &  50.9 \\
    Shona$^{\star}$     & \texttt{sna} &   198 & 14,109 & 1,727 & 1,749 & 17,585 &  99.2 \\
    Sidama$^{\dagger}$  & \texttt{sid} &   395 & 38,752 & 3,091 & 3,561 & 45,404 & 223.0 \\
    Soga$^{\star}$      & \texttt{sog} &   168 &  6,219 &   841 &   817 &  7,877 &  50.3 \\
    Tigrinya$^{\dagger}$& \texttt{tir} &   385 & 40,856 & 4,425 & 5,034 & 50,315 & 217.0 \\
    Wolaytta$^{\dagger}$& \texttt{wal} &   340 & 41,833 & 2,444 & 3,014 & 47,291 & 217.0 \\
    \midrule
    \textbf{Total} & --- & \textbf{5,021} & \textbf{362,123} & \textbf{36,294} & \textbf{39,812} & \textbf{438,229} & \textbf{2,339.7} \\
    \bottomrule
  \end{tabular}
\end{table}

\begin{table*}[ht!]
\centering
\caption{Word Error Rate (WER, decimal; lower is better) on the speaker-disjoint \textsc{WAXAL} test sets (\texttt{v1} for 6 Voice Donor languages ($\dagger$) with $0\%$ train--test speaker overlap; \texttt{v2} for 13 re-split Makerere and Clear Global languages ($\star$)) before (\textbf{Zero-Shot}) and after (\textbf{Fine-Tuned}) monolingual fine-tuning. Fine-tuned entries report $\text{Mean} \pm \text{95\% CI}$ across $N \ge 3$ independent training runs per language ($N \in [3, 9]$). Bold values indicate the better result between Zero-Shot and Fine-Tuned within each model pair.}
\label{tab:asr_results}
\resizebox{\textwidth}{!}{
\begin{tabular}{ll|cc|cc|cc}
\toprule
& & \multicolumn{2}{c|}{\textbf{Gemma-3n-E2B-it}} & \multicolumn{2}{c|}{\textbf{Whisper-large-v3}} & \multicolumn{2}{c}{\textbf{MMS-1B-all}} \\
\textbf{Language} & \textbf{ISO} & \textbf{Zero-Shot} & \textbf{Fine-Tuned ($\pm$95\% CI)} & \textbf{Zero-Shot} & \textbf{Fine-Tuned ($\pm$95\% CI)} & \textbf{Zero-Shot} & \textbf{Fine-Tuned ($\pm$95\% CI)} \\
\midrule
Acholi$^{\star}$ & \texttt{ach} & 0.9398 & \textbf{0.4175 $\pm$ 0.0052} & 1.0552 & \textbf{0.3853 $\pm$ 0.0146} & 0.5059 & \textbf{0.4158 $\pm$ 0.0445} \\
Akan$^{\dagger}$ & \texttt{aka} & 0.9295 & \textbf{0.3533 $\pm$ 0.0385} & 0.9917 & \textbf{0.3477 $\pm$ 0.0254} & 0.5699 & \textbf{0.3900 $\pm$ 0.0074} \\
Amharic$^{\dagger}$ & \texttt{amh} & 0.8568 & \textbf{0.4120 $\pm$ 0.0903} & 1.5240 & \textbf{0.9314 $\pm$ 0.0365} & 0.5715 & \textbf{0.4761 $\pm$ 0.0388} \\
Dagbani$^{\star}$ & \texttt{dag} & 0.9784 & \textbf{0.4337 $\pm$ 0.0197} & 1.2057 & \textbf{0.4417 $\pm$ 0.0475} & --- & \textbf{0.4754 $\pm$ 0.0027} \\
Dagaare$^{\star}$ & \texttt{dga} & 0.9502 & \textbf{0.4456 $\pm$ 0.0171} & 0.9894 & \textbf{0.4066 $\pm$ 0.0064} & 0.5148 & \textbf{0.4133 $\pm$ 0.0042} \\
Ewe$^{\star}$ & \texttt{ewe} & 1.0371 & \textbf{0.4165 $\pm$ 0.0184} & 1.0063 & \textbf{0.4595 $\pm$ 0.0201} & 0.7032 & \textbf{0.4720 $\pm$ 0.0776} \\
Fula$^{\star}$ & \texttt{ful} & 0.9479 & \textbf{0.5625 $\pm$ 0.0179} & 1.2235 & \textbf{0.4688 $\pm$ 0.0144} & 0.5791 & \textbf{0.4408 $\pm$ 0.0066} \\
Ikposo$^{\star}$ & \texttt{kpo} & 1.0430 & \textbf{0.8074 $\pm$ 0.0086} & 1.0937 & \textbf{0.7780 $\pm$ 0.0204} & --- & \textbf{0.7574 $\pm$ 0.0384} \\
Lingala$^{\star}$ & \texttt{lin} & 0.7066 & \textbf{0.4093 $\pm$ 0.0192} & 1.0328 & \textbf{0.3348 $\pm$ 0.0148} & \textbf{0.2885} & 0.3325 $\pm$ 0.0422 \\
Luganda$^{\star}$ & \texttt{lug} & 0.8995 & \textbf{0.2655 $\pm$ 0.0032} & 1.0790 & \textbf{0.2428 $\pm$ 0.0100} & 0.3530 & \textbf{0.1885 $\pm$ 0.0050} \\
Masaaba$^{\star}$ & \texttt{mas} & 1.0641 & \textbf{0.6225 $\pm$ 0.0125} & 1.3189 & \textbf{0.5602 $\pm$ 0.0088} & 0.7519 & \textbf{0.5537 $\pm$ 0.0106} \\
Malagasy$^{\star}$ & \texttt{mlg} & 0.9405 & \textbf{0.2281 $\pm$ 0.0069} & 1.3415 & \textbf{1.0560 $\pm$ 0.2094} & 0.4166 & \textbf{0.3314 $\pm$ 0.0200} \\
Runyankole$^{\star}$ & \texttt{nyn} & 1.0009 & \textbf{0.3773 $\pm$ 0.0229} & 1.1354 & \textbf{0.3612 $\pm$ 0.0057} & 0.5154 & \textbf{0.3951 $\pm$ 0.0475} \\
Oromo$^{\dagger}$ & \texttt{orm} & 0.9426 & \textbf{0.3096 $\pm$ 0.0199} & 1.3016 & \textbf{0.2689 $\pm$ 0.0122} & 0.3313 & \textbf{0.3065 $\pm$ 0.0030} \\
Shona$^{\star}$ & \texttt{sna} & 0.8250 & \textbf{0.3381 $\pm$ 0.0139} & 1.2138 & \textbf{0.2897 $\pm$ 0.0153} & 0.3662 & \textbf{0.3049 $\pm$ 0.0128} \\
Sidama$^{\dagger}$ & \texttt{sid} & 1.1209 & \textbf{0.3789 $\pm$ 0.0408} & 1.4415 & \textbf{0.5148 $\pm$ 0.1800} & \textbf{0.3423} & 0.3431 $\pm$ 0.0018 \\
Soga$^{\star}$ & \texttt{sog} & 0.9326 & \textbf{0.4830 $\pm$ 0.0015} & 1.1165 & \textbf{0.4711 $\pm$ 0.0133} & 0.5738 & \textbf{0.3975 $\pm$ 0.0192} \\
Tigrinya$^{\dagger}$ & \texttt{tir} & 1.3150 & \textbf{0.5561 $\pm$ 0.0764} & 1.6837 & \textbf{0.7727 $\pm$ 0.0142} & 0.6999 & \textbf{0.6150 $\pm$ 0.0173} \\
Wolaytta$^{\dagger}$ & \texttt{wal} & 1.2005 & \textbf{0.4337 $\pm$ 0.0058} & 1.5208 & \textbf{0.6402 $\pm$ 0.3376} & 0.6128 & \textbf{0.4338 $\pm$ 0.0055} \\
\midrule
\textbf{Macro-Average} & --- & 0.9806 & \textbf{0.4342 $\pm$ 0.0231} & 1.2250 & \textbf{0.5122 $\pm$ 0.0530} & 0.5115$^{\ddagger}$ & \textbf{0.4233 $\pm$ 0.0213} \\
\bottomrule
\multicolumn{8}{l}{\footnotesize $^{\ddagger}$Zero-shot macro-average for \texttt{MMS-1B-all} is computed over the 17 languages with pretrained adapters (\texttt{dag} and \texttt{kpo} have no zero-shot adapter).}
\end{tabular}
}
\end{table*}

\begin{table*}[ht!]
\centering
\caption{Character Error Rate (CER, decimal; lower is better) and number of independent training runs ($N$) per language on the speaker-disjoint \textsc{WAXAL} test sets before (\textbf{Zero-Shot}) and after (\textbf{Fine-Tuned}) monolingual fine-tuning. Fine-tuned entries report $\text{Mean} \pm \text{95\% CI}$ over $N$ runs.}
\label{tab:waxal_cer_appendix}
\resizebox{\textwidth}{!}{
\begin{tabular}{ll|ccc|ccc|ccc}
\toprule
& & \multicolumn{3}{c|}{\textbf{Gemma-3n-E2B-it}} & \multicolumn{3}{c|}{\textbf{Whisper-large-v3}} & \multicolumn{3}{c}{\textbf{MMS-1B-all}} \\
\textbf{Language} & \textbf{ISO} & \textbf{ZS} & \textbf{Fine-Tuned ($\pm$95\% CI)} & $N$ & \textbf{ZS} & \textbf{Fine-Tuned ($\pm$95\% CI)} & $N$ & \textbf{ZS} & \textbf{Fine-Tuned ($\pm$95\% CI)} & $N$ \\
\midrule
Acholi$^{\star}$ & \texttt{ach} & 0.4656 & \textbf{0.2138 $\pm$ 0.0078} & 3 & 0.6487 & \textbf{0.1783 $\pm$ 0.0118} & 3 & 0.1848 & \textbf{0.1337 $\pm$ 0.0160} & 6 \\
Akan$^{\dagger}$ & \texttt{aka} & 0.4380 & \textbf{0.1422 $\pm$ 0.0364} & 3 & 0.4543 & \textbf{0.1185 $\pm$ 0.0058} & 3 & 0.1914 & \textbf{0.1222 $\pm$ 0.0019} & 3 \\
Amharic$^{\dagger}$ & \texttt{amh} & 0.5973 & \textbf{0.2113 $\pm$ 0.1159} & 6 & 1.5126 & \textbf{0.7286 $\pm$ 0.0077} & 3 & 0.2003 & \textbf{0.1614 $\pm$ 0.0146} & 3 \\
Dagbani$^{\star}$ & \texttt{dag} & 0.5082 & \textbf{0.1677 $\pm$ 0.0143} & 3 & 0.7623 & \textbf{0.1576 $\pm$ 0.0237} & 3 & --- & \textbf{0.1555 $\pm$ 0.0008} & 3 \\
Dagaare$^{\star}$ & \texttt{dga} & 0.5017 & \textbf{0.1988 $\pm$ 0.0177} & 3 & 0.5478 & \textbf{0.1567 $\pm$ 0.0032} & 3 & 0.1915 & \textbf{0.1495 $\pm$ 0.0019} & 3 \\
Ewe$^{\star}$ & \texttt{ewe} & 0.5487 & \textbf{0.1554 $\pm$ 0.0133} & 3 & 0.4985 & \textbf{0.1612 $\pm$ 0.0064} & 3 & 0.2467 & \textbf{0.1471 $\pm$ 0.0206} & 3 \\
Fula$^{\star}$ & \texttt{ful} & 0.3951 & \textbf{0.2273 $\pm$ 0.0514} & 3 & 0.7150 & \textbf{0.1336 $\pm$ 0.0036} & 3 & 0.1773 & \textbf{0.1198 $\pm$ 0.0014} & 3 \\
Ikposo$^{\star}$ & \texttt{kpo} & 0.5158 & \textbf{0.3082 $\pm$ 0.0209} & 3 & 0.5495 & \textbf{0.2766 $\pm$ 0.0030} & 3 & --- & \textbf{0.2570 $\pm$ 0.0219} & 3 \\
Lingala$^{\star}$ & \texttt{lin} & 0.2628 & \textbf{0.1835 $\pm$ 0.0114} & 3 & 0.5068 & \textbf{0.1120 $\pm$ 0.0051} & 3 & \textbf{0.1002} & 0.1101 $\pm$ 0.0141 & 6 \\
Luganda$^{\star}$ & \texttt{lug} & 0.3128 & \textbf{0.1187 $\pm$ 0.0044} & 3 & 0.4285 & \textbf{0.0974 $\pm$ 0.0018} & 3 & 0.0759 & \textbf{0.0376 $\pm$ 0.0016} & 3 \\
Masaaba$^{\star}$ & \texttt{mas} & 0.3681 & \textbf{0.1903 $\pm$ 0.0129} & 3 & 0.5128 & \textbf{0.1366 $\pm$ 0.0020} & 3 & 0.1817 & \textbf{0.1195 $\pm$ 0.0041} & 3 \\
Malagasy$^{\star}$ & \texttt{mlg} & 0.4954 & \textbf{0.0971 $\pm$ 0.0055} & 3 & 0.7738 & \textbf{0.5027 $\pm$ 0.1107} & 9 & 0.1033 & \textbf{0.0772 $\pm$ 0.0048} & 3 \\
Runyankole$^{\star}$ & \texttt{nyn} & 0.3880 & \textbf{0.1288 $\pm$ 0.0197} & 3 & 0.4426 & \textbf{0.0981 $\pm$ 0.0029} & 3 & 0.1197 & \textbf{0.0896 $\pm$ 0.0103} & 3 \\
Oromo$^{\dagger}$ & \texttt{orm} & 0.3545 & \textbf{0.0980 $\pm$ 0.0219} & 3 & 0.7848 & \textbf{0.0661 $\pm$ 0.0017} & 3 & 0.0769 & \textbf{0.0705 $\pm$ 0.0005} & 3 \\
Shona$^{\star}$ & \texttt{sna} & 0.2544 & \textbf{0.0810 $\pm$ 0.0068} & 3 & 0.3808 & \textbf{0.0522 $\pm$ 0.0024} & 3 & 0.0680 & \textbf{0.0512 $\pm$ 0.0015} & 3 \\
Sidama$^{\dagger}$ & \texttt{sid} & 0.5825 & \textbf{0.1044 $\pm$ 0.0120} & 3 & 0.7721 & \textbf{0.2012 $\pm$ 0.1262} & 6 & 0.0799 & \textbf{0.0778 $\pm$ 0.0003} & 3 \\
Soga$^{\star}$ & \texttt{sog} & 0.3171 & \textbf{0.1381 $\pm$ 0.0074} & 3 & 0.4552 & \textbf{0.1151 $\pm$ 0.0009} & 3 & 0.1190 & \textbf{0.0750 $\pm$ 0.0038} & 3 \\
Tigrinya$^{\dagger}$ & \texttt{tir} & 1.1135 & \textbf{0.2882 $\pm$ 0.0687} & 3 & 1.7035 & \textbf{0.4626 $\pm$ 0.0050} & 3 & 0.2965 & \textbf{0.2411 $\pm$ 0.0088} & 3 \\
Wolaytta$^{\dagger}$ & \texttt{wal} & 0.6673 & \textbf{0.1353 $\pm$ 0.0077} & 3 & 0.8189 & \textbf{0.2420 $\pm$ 0.1862} & 6 & 0.1633 & \textbf{0.1212 $\pm$ 0.0012} & 3 \\
\midrule
\textbf{Macro-Average} & --- & 0.4783 & \textbf{0.1678 $\pm$ 0.0240} & 60 & 0.6983 & \textbf{0.2104 $\pm$ 0.0268} & 69 & 0.1515$^{\ddagger}$ & \textbf{0.1219 $\pm$ 0.0068} & 63 \\
\bottomrule
\multicolumn{11}{l}{\footnotesize $^{\ddagger}$Zero-shot macro-average for \texttt{MMS-1B-all} is computed over the 17 languages with pretrained adapters ($N$ column bottom row shows total runs per model: $60 + 69 + 63 = 192$ runs).}
\end{tabular}
}
\end{table*}

\subsection{Speaker-Leakage Ablation: Impact of Speaker Disjointness}
\label{app:leakage_ablation}

To isolate the impact of speaker overlap, we compare fine-tuned performance on the initial utterance-random splits (\texttt{v1}) against the strictly speaker-disjoint splits (\texttt{v2}) across all 13 re-split Makerere and Clear Global languages (Table~\ref{tab:leakage_ablation}). Removing speaker overlap entirely increases average fine-tuned WER by only $+0.0317$ ($+3.17$ percentage points, from $0.4150$ to $0.4467$) for Gemma 3n 2B, $+0.0200$ ($+2.00$ pp, from $0.4612$ to $0.4812$) for Whisper-Large-v3, and $+0.0213$ ($+2.13$ pp, from $0.3986$ to $0.4199$) for MMS-1B-all. Over $\mathbf{94\%}$ of the total WER reduction achieved over zero-shot baselines is retained under strict speaker disjointness, confirming that models trained on WAXAL learn generalizable acoustic-phonetic representations rather than speaker-specific traits.

\begin{table}[ht!]
\centering
\footnotesize
\caption{Speaker-Leakage Ablation (\texttt{v1} Speaker-Overlapping vs.\ \texttt{v2} Speaker-Disjoint Splits across the 13 re-split Makerere and Clear Global languages). Over $94\%$ of the WER reduction over Zero-Shot is preserved after removing all speaker overlap.}
\label{tab:leakage_ablation}
\begin{tabular}{lcccc}
\toprule
\textbf{Model} & \textbf{ZS WER (\texttt{v2})} & \textbf{FT WER (\texttt{v1})} & \textbf{FT WER (\texttt{v2})} & \textbf{$\Delta$ (\texttt{v2} $-$ \texttt{v1})} \\
\midrule
\texttt{Gemma-3n-E2B-it} & 0.9435 & 0.4150 & 0.4467 & $+0.0317$ ($+3.17$ pp) \\
\texttt{Whisper-large-v3} & 1.1394 & 0.4612 & 0.4812 & $+0.0200$ ($+2.00$ pp) \\
\texttt{MMS-1B-all} & 0.5146 & 0.3986 & 0.4199 & $+0.0213$ ($+2.13$ pp) \\
\bottomrule
\end{tabular}
\end{table}

\begin{table*}[ht!]
\centering
\caption{Out-of-Distribution (OOD) Generalization Word Error Rate (WER) and Character Error Rate (CER) across three independent benchmarks: \textbf{FLEURS v2} (6 languages), \textbf{Mozilla Common Voice} (\texttt{v1 + lg}, 4 languages), and \textbf{Sunbird SALT} (3 East African languages). Bold indicates better WER between Zero-Shot (\textbf{ZS}) and \textsc{WAXAL}-Fine-Tuned (\textbf{FT}).}
\label{tab:ood_all_appendix}
\label{tab:fleurs_results}
\resizebox{\textwidth}{!}{
\begin{tabular}{ll|cc|cc|cc}
\toprule
& & \multicolumn{2}{c|}{\textbf{Gemma-3n-E2B-it (WER / CER)}} & \multicolumn{2}{c|}{\textbf{Whisper-large-v3 (WER / CER)}} & \multicolumn{2}{c}{\textbf{MMS-1B-all (WER / CER)}} \\
\textbf{Benchmark} & \textbf{Language} & \textbf{Zero-Shot} & \textbf{Fine-Tuned} & \textbf{Zero-Shot} & \textbf{Fine-Tuned} & \textbf{Zero-Shot} & \textbf{Fine-Tuned} \\
\midrule
\multirow{7}{*}{\textbf{FLEURS v2}} & Luganda (\texttt{lug}) & 0.8810 / 0.2304 & \textbf{0.6427 / 0.2039} & 1.0349 / 0.2795 & \textbf{0.5432 / 0.1448} & \textbf{0.3848 / 0.0756} & 0.4594 / 0.1107 \\
 & Lingala (\texttt{lin}) & 0.6222 / 0.2154 & \textbf{0.5833 / 0.3326} & 0.8306 / 0.3036 & \textbf{0.2771 / 0.0910} & \textbf{0.1471 / 0.0420} & 0.1702 / 0.0460 \\
 & Oromo (\texttt{orm}) & 0.9192 / 0.3186 & \textbf{0.7311 / 0.2712} & 1.0118 / 0.4449 & \textbf{0.6252 / 0.1773} & \textbf{0.6437 / 0.1724} & 0.7008 / 0.1877 \\
 & Amharic (\texttt{amh}) & 0.8515 / 0.7144 & \textbf{0.4252 / 0.1768} & 1.3832 / 1.6091 & \textbf{1.1568 / 1.0720} & \textbf{0.2957 / 0.0824} & 0.3819 / 0.1057 \\
 & Fula (\texttt{ful}) & 0.8710 / 0.3475 & \textbf{0.7165 / 0.3458} & 0.9009 / 0.3240 & \textbf{0.5419 / 0.1851} & \textbf{0.4891 / 0.1380} & 0.5007 / 0.1490 \\
 & Shona (\texttt{sna}) & 0.8847 / 0.3095 & \textbf{0.6276 / 0.2701} & 1.1200 / 0.2970 & \textbf{0.3960 / 0.1180} & \textbf{0.2271 / 0.0434} & 0.3302 / 0.0751 \\
 & \textit{Macro-Average (6 Langs)} & \textit{0.8383} / \textit{0.3560} & \textbf{\textit{0.6211} / \textit{0.2667}} & \textit{1.0469} / \textit{0.5430} & \textbf{\textit{0.5900} / \textit{0.2980}} & \textbf{\textit{0.3646} / \textit{0.0923}} & \textit{0.4239} / \textit{0.1124} \\
\midrule
\multirow{5}{*}{\textbf{Common Voice}} & Amharic (\texttt{amh}) & 1.5113 / 1.2765 & \textbf{0.6700 / 0.3475} & 1.6855 / 1.9926 & \textbf{1.3186 / 1.1692} & 0.5351 / 0.1846 & \textbf{0.4846 / 0.1495} \\
 & Dagbani (\texttt{dag}) & 1.4209 / 0.9129 & \textbf{0.9168 / 0.5076} & 1.0231 / 0.6145 & \textbf{0.5985 / 0.1714} & --- & \textbf{0.6404 / 0.1672} \\
 & Luganda (\texttt{lug}) & 1.1892 / 0.4541 & \textbf{0.5729 / 0.1831} & 1.1581 / 0.4416 & \textbf{0.4767 / 0.1129} & --- & --- \\
 & Oromo (\texttt{orm}) & 2.0404 / 1.2209 & \textbf{0.9928 / 0.7489} & 1.2567 / 0.7881 & \textbf{0.3354 / 0.0913} & 0.3859 / 0.0928 & \textbf{0.3472 / 0.0740} \\
 & \textit{Macro-Average (4 Langs)} & \textit{1.5404} / \textit{0.9661} & \textbf{\textit{0.7881} / \textit{0.4467}} & \textit{1.2809} / \textit{0.9592} & \textbf{\textit{0.6823} / \textit{0.3862}} & \textbf{\textit{0.4605} / \textit{0.1387}} & \textit{0.4908} / \textit{0.1303} \\
\midrule
\multirow{4}{*}{\textbf{SALT}} & Acholi (\texttt{ach}) & 0.9744 & \textbf{0.4588} & 1.1238 & \textbf{0.7301} & --- & --- \\
 & Luganda (\texttt{lug}) & 0.9312 & \textbf{0.2987} & 1.1374 & \textbf{0.3830} & --- & --- \\
 & Runyankole (\texttt{nyn}) & 0.9861 & \textbf{0.4617} & 1.1780 & \textbf{0.5460} & --- & --- \\
 & \textit{Macro-Average (3 Langs)} & \textit{0.9639} & \textbf{\textit{0.4064}} & \textit{1.1464} & \textbf{\textit{0.5530}} & --- & --- \\
\bottomrule
\end{tabular}
}
\end{table*}

\begin{table}[ht!]
\centering
\footnotesize
\caption{Bidirectional Cross-Dataset Evaluation and Generalization Gap (\texttt{MMS-1B} Adapter WER across 6 overlapping languages). $\text{MMS}_{A|B}$ denotes \texttt{MMS-1B} with adapters fine-tuned on dataset $A$ and evaluated on dataset $B$; \textbf{Diff} reports the in-domain minus cross-domain WER difference ($\text{MMS}_{B|B} - \text{MMS}_{A|B}$).}
\label{tab:cross_train}
\resizebox{\textwidth}{!}{
\begin{tabular}{l|ccc|ccc}
\toprule
& \multicolumn{3}{c|}{\textbf{Evaluated on \textsc{WAXAL} Test}} & \multicolumn{3}{c}{\textbf{Evaluated on FLEURS Test}} \\
\textbf{Language} & $\mathbf{MMS}_{\text{WAXAL}|\text{WAXAL}}$ & $\mathbf{MMS}_{\text{FLEURS}|\text{WAXAL}}$ & \textbf{Diff} & $\mathbf{MMS}_{\text{FLEURS}|\text{FLEURS}}$ & $\mathbf{MMS}_{\text{WAXAL}|\text{FLEURS}}$ & \textbf{Diff} \\
\midrule
Luganda (\texttt{lug}) & \textbf{0.19} & 0.35 & $-0.16$ & \textbf{0.39} & 0.46 & $-0.07$ \\
Lingala (\texttt{lin}) & 0.33 & \textbf{0.29} & $+0.04$ & \textbf{0.15} & 0.17 & $-0.02$ \\
Oromo (\texttt{orm})   & \textbf{0.31} & 0.33 & $-0.02$ & \textbf{0.64} & 0.70 & $-0.06$ \\
Amharic (\texttt{amh}) & \textbf{0.48} & 0.57 & $-0.09$ & \textbf{0.30} & 0.38 & $-0.08$ \\
Fula (\texttt{ful})    & \textbf{0.44} & 0.58 & $-0.14$ & \textbf{0.49} & 0.50 & $-0.01$ \\
Shona (\texttt{sna})   & \textbf{0.30} & 0.37 & $-0.07$ & \textbf{0.23} & 0.33 & $-0.10$ \\
\midrule
\textit{Macro-Average (6 Langs)} & \textbf{\textit{0.34}} & \textit{0.42} & \textit{$-0.07$} & \textbf{\textit{0.37}} & \textit{0.42} & \textit{$-0.06$} \\
\bottomrule
\end{tabular}
}
\end{table}

\section{LLM-as-judge and Qualitative Examples}
\label{sec:app_llm_judge}
Table~\ref{tab:human_llm_agreement} reports the human validation correlation and error metrics between native speaker annotations and the Gemini 3.1 Pro continuous semantic judge on Akan ($n=50$) and Amharic ($n=49$).
Table~\ref{tab:llm_judge} presents the average scores for each language with the judge using both a discrete and continuous rating scale.
Table~\ref{tab:qual_examples} presents representative examples from the LLM-as-judge evaluation of Gemma 3n 2B fine-tuned predictions for Akan and Amharic, which show large Discrete--Continuous score gaps (+0.105 and +0.114 respectively). Note that the utterance-level LLM-as-judge qualitative examples (Table~\ref{tab:llm_judge}, Table~\ref{tab:qual_examples_akan4}, and Table~\ref{tab:qual_examples}) and the diagnostic error-mode counts in Appendix~\ref{app:whisper_failures} were evaluated on the initial (\texttt{v1}) partition; for Akan and Amharic (both Voice Donor languages), the \texttt{v1} and \texttt{v2} test splits are identical ($0\%$ speaker overlap).

\begin{table}[h!]
\centering
\footnotesize
\caption{Human vs.\ LLM-as-judge (Gemini 3.1 Pro) agreement on the continuous $[0.0, 1.0]$ semantic meaning-preservation scale across randomly sampled test utterances rated independently by native speakers of Akan and Amharic.}
\label{tab:human_llm_agreement}
\begin{tabular}{lcccc}
\toprule
\textbf{Language} & \textbf{Samples ($N$)} & \textbf{Pearson $r$} & \textbf{MAE} & \textbf{Mean Bias ($\text{LLM} - \text{Human}$)} \\
\midrule
Akan (\texttt{aka})    & 50 & 0.74 & 0.13 & $+0.01$ \\
Amharic (\texttt{amh}) & 49 & 0.68 & 0.14 & $+0.05$ \\
\midrule
\textbf{Total / Overall} & \textbf{99} & \textbf{0.71} & \textbf{0.14} & $\mathbf{+0.03}$ \\
\bottomrule
\end{tabular}
\end{table}

\begin{table}[h!]
\centering
\footnotesize
\setlength{\tabcolsep}{4pt}
\caption{LLM-as-judge meaning-preservation scores for Gemma 3n 2B fine-tuned predictions. Judge model: Gemini-3.1-pro. Discrete scores $\in\{0.0, 0.5, 1.0\}$; Continuous scores $\in[0.0, 1.0]$. $\Delta$ = Cont $-$ Disc. Higher is better.}
\label{tab:llm_judge}
\begin{tabular}{lcccc}
\toprule
\textbf{Language} & \textbf{Discrete} & \textbf{Continuous} & $\boldsymbol{\Delta}$ & \textbf{Examples} \\
\midrule
Amharic    & 0.574 & 0.688 & +0.114 & 1,710 \\
Tigrinya   & 0.514 & 0.619 & +0.105 & 2,517 \\
Oromo      & 0.737 & 0.805 & +0.069 & 1,891 \\
Soga       & 0.695 & 0.774 & +0.080 &   408 \\
Wolaytta   & 0.804 & 0.837 & +0.034 & 1,507 \\
Sidama     & 0.775 & 0.816 & +0.041 & 1,780 \\
Masaaba    & 0.639 & 0.705 & +0.066 &   429 \\
Luganda    & 0.699 & 0.796 & +0.097 &   319 \\
Nyankole   & 0.668 & 0.753 & +0.085 &   429 \\
Shona      & 0.784 & 0.848 & +0.064 &   874 \\
Lingala    & 0.652 & 0.707 & +0.054 &   933 \\
Akan       & 0.611 & 0.716 & +0.105 &   761 \\
Ewe        & 0.607 & 0.703 & +0.096 &   945 \\
Dagbani    & 0.665 & 0.752 & +0.087 &   919 \\
Dagaare    & 0.737 & 0.791 & +0.055 &   955 \\
Fulani     & 0.760 & 0.816 & +0.056 & 1,195 \\
Ikposo     & 0.511 & 0.583 & +0.072 &   922 \\
Acholi     & 0.527 & 0.653 & +0.127 &   264 \\
Malagasy   & 0.713 & 0.776 & +0.064 & 1,145 \\
\midrule
\textbf{Macro-Avg / Total} & \textbf{0.667} & \textbf{0.744} & \textbf{+0.077} & \textbf{19,903} \\
\bottomrule
\end{tabular}
\end{table}

\textbf{Qualitative Examples: LLM-as-Judge Scoring.}
\label{subsec:qual_examples}
Table~\ref{tab:qual_examples} presents representative examples from the LLM-as-judge evaluation of Gemma 3n 2B fine-tuned predictions for Akan and Amharic, which show large Discrete--Continuous score gaps (+0.105 and +0.114 respectively). These examples illustrate the three scoring tiers and the corresponding failure modes.

\begingroup
\scriptsize
\setlength{\tabcolsep}{2pt}
\setlength{\LTcapwidth}{\textwidth}
\begin{xltabular}{\textwidth}{c X X X}
\caption{Qualitative examples of Gemma 3n 2B zero-shot and fine-tuned predictions for Akan and Amharic, graded by the LLM-as-judge. Both Discrete (D) and Continuous (C) scores on fine-tuned predictions. WER is Word Error Rate along with Explanations.}\label{tab:qual_examples}\\
\toprule
\textbf{Score} & \textbf{Reference} & \textbf{Zero shot prediction (WER)} & \textbf{Fine tuned prediction (WER)} \\
\midrule
\endfirsthead
\caption[]{Qualitative examples (continued)}\\
\toprule
\textbf{Score} & \textbf{Reference} & \textbf{Zero shot prediction (WER)} & \textbf{Fine tuned prediction (WER)} \\
\midrule
\endhead
\multicolumn{4}{r}{Continued on next page} \\
\bottomrule
\endfoot
\bottomrule
\endlastfoot
\multicolumn{4}{c}{\textbf{Akan}} \\
\midrule
\makecell[t]{D:0.0\\C:0.15} & Nkwadaa bi tete atweredeɛ so wɔ ɛdan bi akyi. Ɛdan no ho ayɛ fi. Bɔtɔ deda wɔn anim. Baabi a wɔwɔ no, nneɛma fi bebree wɔ hɔ. Rɔba tuntum bi si wɔn anim. & --- & Dan bi sisi atwitwa so wɔ bɔn bi akyi. Dan no hi yɛ fi. Bɔtɔ deda ara. Baabi a wɔtete hɔ no nan dun bi ho. Wɔakata ntwerɛni no. (WER: 70\%) \\*
\multicolumn{4}{p{\dimexpr\textwidth-2\tabcolsep\relax}}{\textbf{Explanation:} The reference 'Nkwadaa bi tete atweredeɛ so wɔ ɛdan bi akyi' (Some children are sitting on stairs behind a building) is completely changed in the prediction 'Dan bi sisi atwitwa so wɔ bɔn bi akyi' (Some buildings are sitting on something cut down behind a valley). This is a critical failure changing the subject from 'children' to 'buildings' and the location completely.} \\
\midrule
\makecell[t]{D:0.0\\C:0.1} & Mmaa sɛ wogyina gyina hɔ. Ebi de ne nsa asɔ ne sisi. Ebi ɛde wɔn nsa aka abom. Ebi egyina hɔ ɛde wɔn nsa egu wɔn bo. Ebi nso ɛsere. Ataade a ɛhyɛ ebinom ɛyɛ fitaa. Ebi de yɛ ahahanmono, ɛna ebi so de yɛ kɔkɔɔ. & Enma, bi se fo kenee gya ho. Ebi ti nsa aso wo nsisi. Ebi atie wo nsa akan wo. Ebi nso esere atalia ehu ebini ye fita. Ebi de ahamono ne bi denso yekoko. (WER: 85\%) & Mmaa bi hyɛ adwadipono ho. Ebi de ne nsa asɔ wɔn sisi. Ebi de wɔn nsa aka aboom. Ebi akyi ahokɔsradeɛ nso wɔ wɔn ho. (WER: 71\%) \\*
\multicolumn{4}{p{\dimexpr\textwidth-2\tabcolsep\relax}}{\textbf{Explanation:} The reference describes women standing ('Mmaa sɛ wogyina gyina hɔ')... The prediction fundamentally changes the meaning right from the start. It says 'Mmaa bi hyɛ adwadipono ho' (some women are at a shop/counter), replacing the description of them standing.} \\
\midrule
\makecell[t]{D:0.5\\C:0.85} & Mmayewa mmienu, wɔn mmienu nyinaa hyɛ kyɛ. Kanea asɔ wɔ faako a wɔgyina no, na atosodeɛ sisi baabi a wɔgyina no. Deɛ ɔgyina anim no kuta bɔtɔ. & Ɛ\textethi{ማ }wo ameyinu, omue nyinaa a hyɛkye, kannia a sɔwɔ faa akun wo gyina no. Na atosodie esisi baabi a wo mu gyina no, diee ogyinaa nimu no ekuta bot\texttibetan{◌ོ}. (WER: 85\%) & Mmayewa mmienu wɔ mmienu nyinaa hyɛ kyɛ. Kanea asɔ wɔ faako a ogyina no, na atosodeɛ sisi baabi a wɔgyina no. Deɛ ɔgyina anim no kuta bɔtɔ. (WER: 7\%) \\*
\multicolumn{4}{p{\dimexpr\textwidth-2\tabcolsep\relax}}{\textbf{Explanation:} The ZS prediction contains severe phonetic errors and cross-script hallucinations (including Ethiopic and Tibetan). The FT prediction successfully reconstructs the vocabulary to match the reference but incorrectly substitutes the plural pronoun subject prefix in ``wɔgyina'' with the singular ``o'' in ``ogyina'', leading to a partial score.} \\
\midrule
\makecell[t]{D:0.5\\C:0.5} & Adɔfofoɔ ne amanfoɔ a wɔahyia mu a wɔrehwɛ sini. Abusua baako a wɔyɛ nnan rehwɛ wɔn sini, ɛna amanfoɔ pii nso so a wɔahyia mu wɔ sinitwabea, wɔn nso so rehwɛ wɔn sini. & A dafo fo, eny aman fo a, won ehyie mu a won hweene. Ebusia baa h\texttibetan{◌ོ}, a won yɛ enai ehɛ won sinie. En aman fo pi isu so a ehɛ mu ewo sinie trabia wonzo se ehɛ won sina? (WER: 103\%) & Adɔfofoɔ ne amanfoɔ a wɔahyia mu a wɔrehwɛ sini. Abusua baako a wɔyɛ nnan rehyɛ wɔn sini. Ɛnna amanfoɔ pii nso so ahyia mu wɔ sini. N'atwa beaeɛ a wɔn nso so rehyɛ wɔn sini. (WER: 34\%) \\*
\multicolumn{4}{p{\dimexpr\textwidth-2\tabcolsep\relax}}{\textbf{Explanation:} The ZS prediction is heavily fragmented with critical phonetic errors and Tibetan script hallucinations. The FT model successfully reconstructs the sentence structure and core meaning, though it introduces minor lexical substitutions (e.g., ``rehyɛ'' instead of ``rehwɛ'') that slightly alter secondary details.} \\
\midrule
\makecell[t]{D:1.0\\C:1.0} & Nnipa bebree wɔ ha. Nipa no bi rebɔ nhwene. Nipa no bi hyɛ fugu. Nipa no bi resa. Nnipa no bi nso gyina hɔ rehwɛ. Nnua bebree gu wɔn akyi. & Epa bebrewa. Nipa ne bi ɛbɔnshwini. Nipa ne bi shɛfugu. Nipa ne bi ɛsɛ. Nipa ne bi ensu gyina oɔ ɛhɛ. Enia bebreɛgu wo omaiki. (WER: 70\%) & Nnipa bebree wɔ ha. Nnipa no bi rebɔ nhwene. Nnipa no bi hyɛ fugu. Nnipa no bi resa. Nnipa no bi nso gyina hɔ rehwɛ. Nnua bebree gu wɔn akyi. (WER: 37\%) \\*
\multicolumn{4}{p{\dimexpr\textwidth-2\tabcolsep\relax}}{\textbf{Explanation:} The ZS model struggles heavily with poor phonetic spelling and word fragmentation. The FT model resolves these entirely, generating a nearly flawless transcription where the only difference is the benign pluralization of "Nipa" to "Nnipa".} \\
\midrule
\makecell[t]{D:1.0\\C:0.98} & Mmaa bebree gyina hɔ atwa poma ho ahyia, wɔasosɔ wɔn nsa na wɔapagya wɔn nsa akɔ soro tenten. Binom resere. & Emba Bibre, \texttelugu{ఈజినా హో ఎటి}\textethi{ያ አፑማ ሆም ኤሱ ሶ ሆም ሰነ ና ቬፒጃ ሆም ኢንሳኖ አኮስሮ ቴንቴ ኤቢኑ ሞ ኤስ ረ} (WER: 115\%) & Mmaa bebree gyina hɔ atwa poma ho ahyia. Wɔasosɔ wɔn nsa na wɔapagya wɔn nsa no akɔ soro tenten. Binom resere. (WER: 10\%) \\*
\multicolumn{4}{p{\dimexpr\textwidth-2\tabcolsep\relax}}{\textbf{Explanation:} The ZS model severely hallucinates a mix of Telugu and Amharic characters. The FT model completely eliminates this cross-lingual hallucination, producing a near-perfect transcription that differs from the reference only by benign punctuation changes.} \\
\midrule
\multicolumn{4}{c}{\textbf{Amharic}} \\
\midrule
\makecell[t]{D:0.0\\C:0.2} & \textethi{ጠብታ መስኖ የከፍተኛ የውሃ ብቃት ያለው የመስኖ ዘዴ ነው።ውሃን በቧንቧዎች አማካኝነት በጠብታ መልክ በቀጥታ ወደ ሰብሉ ስር የማዳረስ ሂደት ነው። ዉሃን በእጅጉ ከመቆጠብም በላይ አረም እና የበሽታ ክስተትን ተፅዕኖ ይቀንሳል።} & \textethi{አብተሀመስ ነው። የከፍተኛ የውሃ አብቀት ያለው ዘመናዊ የመስኖች ዘዴ ነው። ውሃን በባምባዎች አማካኝነት በጠብታ መልክ በቀጥታ ወደ }\texthebrew{ישבול }\textethi{ስርት የማዳረ ሲ}de \textethi{ነው። ውሃን በእጅጉ ከመቀተው }\textcyrillic{Вме }\textethi{ላይ የ }år \textethi{ከሆነ} \ldots (WER: 77\%) & \textethi{እቡካተ ብዙሃን ከፋት ያለው ዘመናዊ የመስኖ ዘዴ ነው ውሃን በቧንቧዎች አማካኝነት ገጠታቸው ወደ ሰብሉ ስር የማዳረስ ዘዴ ነው ውሃን በእጅጉ ከመቆጠብ በላይ ለዓረም ኃይል ተጭኖ ይቀንሳል} (WER: 70\%) \\*
\multicolumn{4}{p{\dimexpr\textwidth-2\tabcolsep\relax}}{\textbf{Explanation:} The ZS model hallucinates foreign scripts (Hebrew, Latin, and Cyrillic) and drops significant text. While FT stays within the correct script and improves fluency, it completely hallucinates the subject (replacing "drip irrigation" \textethi{ጠብታ መስኖ} with unrelated words \textethi{እቡካተ ብዙሃን}) and alters key technical terms, failing to convey the core meaning.} \\
\midrule
\makecell[t]{D:0.0\\C:0.2} & \textethi{የምግብ ማምረቻዎች ኩባንያ እነዚህ ፉድ ፕሮሰሲንግ ካምፓኒ የምንለው እንደ አንድ የምግብ ማቀነባበሪያ ፋብሪካዎች ለበጎ አድራጎት ስራ ተብሎ የተወሰኑ ምርቶታቸውን በነፃ ወይም በቅናሽ ይሰጣል።} & \textethi{የመግበ }\texthebrew{רי}\textethi{ናም ረቻዎች ኩባንያዎች ድጋፍ እነዚህ }\texthebrew{הפ}\textethi{ሮሰሲንግ ከ}ampany support \textethi{የምንለው እንደ አንድ የመግምት ማቀናበር የፋብሪካዎች ለበግ አደርጉት ስርዓተው ነው የተወሰኑ ምርቶቻቸውን በንሳ ወይም በከፍተኛ ዋጋ ይሰጣሉ።} (WER: 86\%) & \textethi{የምግብ ማምረቻ ኩባንያዎች ድጋፍ እነዚህ ፍልሰሶች ካምፓኒ የሱ ምርቶች ይካን ፓኒ ያሀፈ ያሬ ፈብሪካዎች ለበጉ አድራጎት ስራ ተብሎ የተወሰኑ ምርቶቻቸውን በእንፃዊ ወይም በቅናሽ ዋጋ} (WER: 70\%) \\*
\multicolumn{4}{p{\dimexpr\textwidth-2\tabcolsep\relax}}{\textbf{Explanation:} The ZS output mixes Hebrew and Latin characters with phonetic approximations. The FT model corrects the script and captures the beginning accurately, but hallucinates heavily in the middle and drops the main verb at the end, resulting in a total loss of the sentence's primary action.} \\
\midrule
\makecell[t]{D:0.5\\C:0.9} & \textethi{ሶስት ወታደራዊ ልብስ የለበሱ ወንዶች ደረቅ መሬት ላይ በአረንጓዴ የደን አካባቢ ቆመዋል። መካከለኛው ሰው ነጭና አረንጓዴ ሻሽ አንገቱ ላይ አድርጓል፤ በቀኝ በኩል ያለው ደግሞ አገዳ ይዟል። እነዚህ ወታደሮች የሀገርን ነፃነት ለመጠበቅ ተሰልፈዋል።} & \textethi{ሶስቱ ተደረዩዎች የለበሰው ወንዶች }\textcyrillic{Даже, }\textethi{መሬት ላይ ባረንጓዴ }\texthebrew{הי}\textethi{ደና አካባቢ }\textcyrillic{Комват}\textethi{።} \textethi{አካለኛው ሰው ሆነጪ እናረንጓዴ }\texttelugu{శా}\textethi{ሻን ቀዶል ያድርጉዋል። በ }kraju \textethi{ሁሉም አለው ደግሞ አገዳ ይዟል። እነዚህ ወታደሮች ያገረነትን ለመጠበቅ ተሰልፈዋል።} (WER: 66\%) & \textethi{ሦስት ወታደራዊ ልብስ የለበሱ ወንዶች ደል መሬት ላይ በአረንጓዴ የደን አካባቢ ቆመዋል። መካከለኛው ሰው ነጭ እና አረንጓዴ ሻሽ አንገቱ ላይ አድርጓል፤ በቀኝ በኩል ያለው ደግሞ አገዳ ይዟል። እነዚህ ወታደሮች የአገርን ነፃነት ለመጠበቅ ተሰልፈዋል።} (WER: 16\%) \\*
\multicolumn{4}{p{\dimexpr\textwidth-2\tabcolsep\relax}}{\textbf{Explanation:} The ZS prediction is heavily corrupted by cross-script hallucinations across four different writing systems (Latin, Hebrew, Cyrillic, Telugu). The FT model eliminates all foreign scripts, yielding a clean Amharic prediction. It scores 0.5 because it uses minor, valid synonyms and spelling variations (e.g., replacing \textethi{ደረቅ} [dry] with \textethi{ደል}) rather than an exact reference match.} \\
\midrule
\makecell[t]{D:0.5\\C:0.8} & \textethi{አምስት ሴቶች ነጭ የባህል ልብስ እና ብርቱካናማ ዶቃዎች አንገት ላይ አድርገው በደስታ ፈገግታ ይታያሉ። ሴቶቹ ጥቁር ፀጉር ያላቸው ሲሆን ከበስተጀርባው አረንጓዴ ገጽታ አለ።} & \textethi{አምስት ሴቶች ነጭ በልብስና ብ}\textmalayalam{\char"25CC ൃത്ത}\textethi{ካን ማህዶ }\textgreek{Κα}\textethi{ዎች አንግት ላይ አድረክ}ko \textethi{በደስታ በፍቅድት ይታዳሉ። ሴቶቹ ትቁር ሰ}\textdevanagari{गौर }\textethi{ያላቸው ሲሆን ከስነጽ}er\textethi{ባቸው አረንጓዴ ክጽታ አለ።} (WER: 57\%) & \textethi{አምስት ሴቶች ነጭ የባህል ልብስ እና ብርቱካናማ ዶቃዎች አንገት ላይ አድርገው በደስታ በፈገግታ ይታያሉ። ሴቶቹ ጥቁር ፀጉር ያላቸው ሲሆን፣ ከሥርባቸው አረንጓዴ ገጽታ አለ።} (WER: 13\%) \\*
\multicolumn{4}{p{\dimexpr\textwidth-2\tabcolsep\relax}}{\textbf{Explanation:} The ZS model incorrectly injects Greek, Devanagari, and Malayalam characters. The FT model perfectly corrects the orthography and captures almost all details, but incorrectly substitutes "behind them" with "under them" (\textethi{ከሥርባቸው}), slightly altering the spatial meaning.} \\
\midrule
\makecell[t]{D:1.0\\C:1.0} & \textethi{ፈገግታ ያላት ወጣት ሴት በመደብሯ ውስጥ ቆማለች። ሱቁ የመዋቢያ ዕቃዎች፣ ሻምፖዎችና የልደት ጌጣጌጦች የተደረደሩበት ነው። ይህም በአዲስ አበባ ወይም በሌላ የኢትዮጵያ ከተማ የሚገኝ አነስተኛ የግል ንግድ ነው።} & \textethi{ተሳገት ያ }\textsinhala{ලද} \textethi{ወጣ ሲህ በመደብረ አወስኮ መለጨት ስኮ የ መዋቢያ ቀዎች }\texthebrew{שא}\textethi{ምፖች እና የልደት ጌታጌጦች }\texthebrew{הי}\textethi{ተደረጉበት ነው። ይህም በአዲስ አበባ ወይም በሌላ የኢትዮጵያ ከተማ}\textdevanagari{एं }\textethi{ሚገኙ }\texthebrew{הנ}\textethi{ስተኛ የክልል ንግድ ነው።} (WER: 74\%) & \textethi{ፈገግታ ያላት ወጣት ሴት በመደብሯ ውስጥ ቆማለች። ሱቁ የመዋቢያ ዕቃዎች፣ ሻምፖዎች እና የልደት ጌጣጌጦች የተደረደሩበት ነው። ይህም በአዲስ አበባ ወይም በሌላ የኢትዮጵያ ከተማ የሚገኝ አነስተኛ የግል ንግድ ነው።} (WER: 7\%) \\*
\multicolumn{4}{p{\dimexpr\textwidth-2\tabcolsep\relax}}{\textbf{Explanation:} The ZS prediction contains multiple phonetic errors and hallucinated Hebrew characters. The FT model generates an output that perfectly preserves the meaning, with the only discrepancy being a completely benign expansion of a conjunction (\textethi{እና} instead of the suffix \textethi{ና}).} \\
\midrule
\makecell[t]{D:1.0\\C:1.0} & \textethi{አራት ጥቁር ባለሁለት ጫፍ የቅንድብ መሳያ እርሳሶች ቀርበዋል። እያንዳንዱ መሳያ በአንድ በኩል ብሩሽ ሲኖረው በሌላኛው በኩል ደግሞ እርሳስ አለው። በሁሉም ላይ በነጭ ቀለም እና በእንግሊዝኛ ፊደላት የተፃፈ ፅሁፍ አለ።} & \textethi{አራት }\textbengali{চক্ষ}\textthaana{\char"25CC ު}\textthaana{ރު }\textethi{ባለ ሁለት }\textcyrillic{Щ}ekend\texthebrew{מס }\textgreek{Α}\textethi{ርሳሶች ቀርበዋል። እያንዳንዱ }\texthebrew{מס}\textethi{ያ በአንድ በኩል ብዙ ሲኖረው በሌላኛው በኩል ደግሞ እርሳሳለው። በሁሉም ላይ በነ }Drive element \textethi{ነው በእንግሊዘኛ }\texttelugu{ఫి}\textethi{ደላት የተጻፈ }\texthebrew{זו}\textethi{።} (WER: 62\%) & \textethi{አራት ጥቁር ባለ ሁለት ጫፍ የቅንድብ መሳያ እርሳሶች ቀርበዋል። እያንዳንዱ መሳያ በአንድ በኩል ብሩሽ ሲኖረው፣ በሌላኛው በኩል ደግሞ እርሳስ አለው። በሁሉም ላይ በነጭ ቀለም እና በእንግሊዝኛ ፊደላት የተጻፈ ጽሑፍ አለ።} (WER: 17\%) \\*
\multicolumn{4}{p{\dimexpr\textwidth-2\tabcolsep\relax}}{\textbf{Explanation:} The ZS model suffers a catastrophic failure, interleaving six different foreign scripts. The FT model successfully outputs the correct text, differing from the reference only through benign orthographic standardizations (e.g., spacing and acceptable spelling variants for \textethi{ጽሑፍ}).} \\
\end{xltabular}
\endgroup

\section{Whisper-Large-v3 Failure Mode Analysis}
\label{app:whisper_failures}

This appendix provides a detailed qualitative analysis of the failure modes exhibited by Whisper-Large-v3, both zero-shot and after fine-tuning on WAXAL, for two languages where aggregate WER does not fully capture the nature of model errors: Amharic and Malagasy.

\subsection{Amharic}
\label{app:whisper_amharic}

Fine-tuned Whisper-Large-v3 achieves a WER of 0.92 on Amharic, significantly worse than Gemma 3n 2B (0.37) and MMS-1b-all (0.48). Qualitative analysis reveals distinct failure modes in both the zero-shot and fine-tuned settings.

\noindent\textbf{Zero-Shot Failures.}
In the zero-shot setting, the dominant failure modes are wrong-script output and hallucinations, as shown below.

\begin{table}[h!]
\centering
\footnotesize
\caption{Whisper-Large-v3 zero-shot failure modes on Amharic (N\,=\,3,420).}
\label{tab:whisper_amh_zeroshot}
\begin{tabular}{lrr}
\toprule
\textbf{Category} & \textbf{Count} & \textbf{Percentage} \\
\midrule
Wrong Script       & 1,737 & 50.8\% \\
Hallucination  & 1,681 & 49.2\% \\
Encoding Corruption         &   2 &  0.1\% \\
\bottomrule
\end{tabular}
\end{table}

\noindent An example of wrong-script output: the reference is Ethiopic text but the model generates Arabic script, e.g., \textit{``\textarabic{ارنغاري كلم يهلو يقول قال قتل\textethi{ና}{\normalfont\itshape\ldots}}''}

\noindent\textbf{Fine-Tuning Failures.}
Fine-tuning resolves several zero-shot failure modes but introduces new ones:

\begin{itemize}[noitemsep,nolistsep,leftmargin=8pt]
    \item \textbf{Wrong-script} outputs are eliminated: 50.8\% $\to$ 0\%.
    \item \textbf{Hallucinations} are substantially reduced: 49.2\% $\to$ 25.8\%.
\end{itemize}

Despite these improvements, the fine-tuned model exhibits persistent failures:

\begin{table}[h!]
\centering
\footnotesize
\caption{Whisper-Large-v3 fine-tuned failure modes on Amharic (N\,=\,3,420).}
\label{tab:whisper_amh_finetuned}
\begin{tabular}{lrr}
\toprule
\textbf{Category} & \textbf{Count} & \textbf{Percentage} \\
\midrule
Encoding Corruption (U+FFFD) & 1,903 & 55.6\% \\
Hallucination (WER $> 0.9$)  &   884 & 25.8\% \\
High WER ($0.5 <$ WER $< 0.9$) & 633 & 18.5\% \\
\bottomrule
\end{tabular}
\end{table}

The dominant failure is the presence of Unicode replacement characters (U+FFFD) in 55.6\% of predictions, indicating that the Ethiopic script is not fully learned during fine-tuning. Even when U+FFFD characters are absent, the character error rate remains high, with words frequently truncated mid-syllable, e.g., \textit{``\textethi{መድኃኒት}''} (medhanit, medicine) $\to$ \textit{``\textethi{መድ}''} (med), \textit{``\textethi{ክፍሎች}''} (kifloch, sections) $\to$ \textit{``\textethi{ክ}''} (ki).

\subsection{Malagasy}
\label{app:whisper_malagasy}

\noindent\textbf{Key Finding.} Fine-tuned Whisper-Large-v3 on Malagasy achieves a WER of 1.10---worse than the zero-shot Gemma 3n 2B (0.94)---and swaps one catastrophic failure mode for another: 76.8\% hallucination in zero-shot becomes $\sim$87\% repetition and stuttering after fine-tuning. Only 14 out of 2,290 samples (0.6\%) are genuinely corrected. The model learns to output Malagasy words but degenerates into repetition loops.

\noindent\textbf{Zero-Shot Failure Distribution.}

\begin{table}[h!]
\centering
\footnotesize
\caption{Whisper-Large-v3 zero-shot failure modes on Malagasy (N\,=\,2,290).}
\label{tab:whisper_mlg_zeroshot}
\begin{tabular}{lrr}
\toprule
\textbf{Failure Mode} & \textbf{Count} & \textbf{\%} \\
\midrule
Hallucination (gibberish / wrong content)  & 1,758 & 76.8\% \\
Wrong Script (Georgian, Arabic, Ethiopic\ldots) & 377 & 16.5\% \\
Short / Truncated                          &    81 &  3.5\% \\
High WER ($>$50\%, correct script)         &    36 &  1.6\% \\
Repetition (stuttering patterns)           &    38 &  1.7\% \\
\bottomrule
\end{tabular}
\end{table}

\noindent\textbf{Fine-Tuned Failure Distribution.}

\begin{table}[h!]
\centering
\footnotesize
\caption{Whisper-Large-v3 fine-tuned failure modes on Malagasy (N\,=\,2,290).}
\label{tab:whisper_mlg_finetuned}
\begin{tabular}{lrr}
\toprule
\textbf{Failure Mode} & \textbf{Count} & \textbf{\%} \\
\midrule
Repetition (stuttering) & 1,982 & 86.6\% \\
Short Output &    196 &  8.6\% \\
High WER ($>$50\%)      &    78 &  3.4\% \\
Hallucination (WER $>$ 0.9)           &    20 &  0.9\% \\
OK (WER $<$ 30\%)       &    14 &  0.6\% \\
\bottomrule
\end{tabular}
\end{table}

\noindent\textbf{Dominant Repetition Patterns.} The fine-tuned model predominantly repeats short fragments of common Malagasy words:

\begin{table}[h!]
\centering
\footnotesize
\caption{Most frequent repetition patterns in Whisper-Large-v3 fine-tuned Malagasy predictions.}
\label{tab:whisper_mlg_repetitions_cleaned}
\begin{tabular}{lr}
\toprule
\textbf{Pattern} & \textbf{Count} \\
\midrule
\texttt{ny ny ny ny\ldots} & 2685 \\
\texttt{i i i i\ldots} & 2685 \\
\texttt{a a a a\ldots} & 2685 \\
\texttt{an an an an\ldots} & 2685 \\
\texttt{I I I I\ldots} &  1115 \\
\texttt{m m m m\ldots} &  1033 \\
\bottomrule
\end{tabular}
\end{table}

\noindent\textbf{Examples of Corrected Predictions.} Despite the overall failure, fine-tuning does resolve the most egregious zero-shot errors in a small fraction of samples:

{
\small\ttfamily
\noindent Example 1:\\
REF: Tovovavy tsara tarehy, fohy volo, miloko mena, miravaka asa tanana maro loko ary manao akanjo mainty\\
ZS:\ \ I am the one who has been given the power to do good. I am the one who has been given the power to d\\
FT:\ \ Tovovavy tsara tarehy, fohy volo miloko mena. Miravaka asa tanana maro loko ary manao akanjo mainty.

\medskip
\noindent Example 2:\\
REF: Tsotsotra kely tsy dia misy haingony loatra ilay firavaka fatao amin'ny tenda izay miloko rano- mbol\\
ZS:\ \ \textarabic{چو چو چا کل، چتیا می سی هنگون لچ، یلی فراوکا فتوام تند، از ام لو کو ران بولا من، کات سمی سم وی دی زی} (wrong script!)\\
FT:\ \ Tsotsra kely tsy dia misy haingony loatra ilay firavaka fatao amin'ny tenda izay miloko ranom-bolame
}

\medskip
\noindent\textbf{Examples of Repetition Failures.}

{
\small\ttfamily
\noindent Example 1:\\
REF:\ \ Andihana lehilahy manao fanamiana mainty ambony ambany, avy nanao baolina miaraka amin'ireo mpanazatra azy anankiroa manao akanjo fotsy sy mainty. Ary hotronin'ireo vehivavy sy zaza ary ireo namana.\\
ZS:\ \ Hello, my name is Lela. I am from Aminah, in the Banyanbani region. I am from the village of Pol. I am from the village of Sarjasa, in the Kuru. I am from the village of Shutsu. I am from the village of Fafsaz, in the Uru-Naman. (English Hallucination)\\
FT:\ \ Andiana lehilah, manao fanamiana aminaina. Manao a a a a a i i i i i

\medskip
\noindent Example 2:\\
REF:\ \ Maro karazany ny zavamaniry noharian'Andriamanitra. Miavaka amin'izy ireny ity voninkazo iray ity, izay tena manitra sady mamerovero, no afaka atao amin'ny sakafo ; ny lokon'ny raviny dia maitso somary marevaka, sy miendrika lavalava.\\
ZS:\ \ Bonjour, je m'appelle Marika Khazan et je suis une amie de mon mari. Je suis une amie de mon fils, qui est en train de se faire un petit déjeuner. Je suis très heureuse de vous voir, et je vous souhaite une bonne journée. (French Hallucination)\\
FT:\ \ \ Maro karazany ny zava maniry, maro kar kar kar kar maro ny ny ny ny ny.
}

\section{LLM-as-Judge Prompts}
\label{app:llm_judge_prompts}

This appendix provides the full system instructions used for the LLM-as-judge meaning-preservation evaluation described in Section~\ref{sec:llm_judge}. We provide the prompt for the discrete scoring rubric in Subsection~\ref{app:discrete_prompt} and the prompt for the continuous scoring rubric in Subsection~\ref{app:continuous_prompt}.

\subsection{Discrete Scoring Prompt}
\label{app:discrete_prompt}

This subsection presents the prompt for the discrete scoring rubric. It guides the judge model to assign scores of 1.0, 0.5, or 0.0 based on whether the core meaning is preserved, partially preserved, or lost.
\vspace{5pt}
\hrule
\begin{small}
\begin{verbatim}
**Role:**
You are an expert linguist and native speaker of various African
languages, tasked with evaluating the performance of an Automatic
Speech Recognition (ASR) system.

**Task:**
You will be provided with the target Language, a Reference
transcription (the actual spoken words), and a Prediction (the
output from the ASR model). Your goal is to analyze the differences
between the Reference and Prediction to determine if the mistakes
made by the ASR model affect the core meaning of the sentence, or
if the text still carries the same meaning despite those mistakes.

**Evaluation Guidelines:**
1. Dialectal and Orthographic Variations (Benign): Differences that
   are merely dialectal, minor spelling variations, or differences
   in tone markings that do not change the understood word in
   context are benign.
2. Loanwords and Phonetic Spellings (Benign): If the model
   transcribes a phonetically similar loanword or correct English
   translation of an English loanword, this is benign.
3. Synonym Hallucinations (Critical): If the prediction replaces a
   word with a synonym that sounds completely different, this is a
   critical failure.
4. Pronoun and Subject Alterations (Critical): Changing pronouns,
   subject markers, or agglutinated prefixes is a critical failure.
5. Insertions and Deletions: Dropping minor filler words is usually
   benign; dropping main verbs or inserting meaning-altering words
   is a critical failure.

**Scoring Rubric:**
* 1.0 (Meaning Preserved): The core meaning is completely intact.
  Any errors are strictly benign.
* 0.5 (Partial Meaning Preserved): The sentence contains a mix of
  benign and critical errors. A native speaker can understand some
  parts, but other parts contain critical errors.
* 0.0 (Meaning Lost): Critical errors overwhelm the sentence.
  Key subjects, verbs, or objects are lost, hallucinated, or
  fundamentally altered.

**Output Format:**
Strictly valid JSON with explanation before score:
{
  "explanation": "...",
  "score": [0.0, 0.5, or 1.0]
}
\end{verbatim}
\end{small}
\hrule

\subsection{Continuous Scoring Prompt}
\label{app:continuous_prompt}

This subsection details the prompt for the continuous scoring rubric. This framework evaluates the Information Transfer Rate on a continuous scale from 0.0 to 1.0, penalizing hallucinations more severely than omissions.
\vspace{5pt}

\hrule
\vspace{5pt}
\begin{small}
\begin{verbatim}
**Role:**
You are an expert linguist and native speaker of various African
languages, tasked with evaluating the performance of an Automatic
Speech Recognition (ASR) system.

**Task:**
You will be provided with the target Language, a Reference
transcription (the actual spoken words), and a Prediction (the
output from the ASR model). Your goal is to analyze the differences
between the Reference and Prediction to determine the Information
Transfer Rate -- how much of the original meaning successfully made
it to a native listener.

**Evaluation Framework:**
1. Dialectal and Orthographic Variations (No Penalty): Differences
   that are merely dialectal, minor spelling variations, or
   differences in tone markings that do not change the understood
   word in context are benign. Phonetically accurate loanwords are
   also completely benign.
2. Primary Intent vs. Secondary Details (Weighted Impact):
   - Primary Intent: The main subject, core action (verb), and
     primary object. Damaging these results in a heavy penalty.
   - Secondary Details: Adverbs, filler words, minor locations, or
     subordinate clauses. Damaging these results in a minor penalty.
3. Omission vs. Hallucination (Error Types):
   - Omission: If the ASR drops a word, the listener loses context
     but is not actively misled. This is a moderate penalty.
   - Hallucination/Misinformation: If the ASR mis-transcribes a
     word into a different valid word that changes the facts, it
     actively lies to the listener. This is a severe penalty.
4. Proportional Impact: Evaluate the damage relative to the
   sentence length.

**Scoring Deciles (0.0 to 1.0):**
* 0.9-1.0 (Perfect/Near Perfect): Core intent and all secondary
  details intact. Variations are benign.
* 0.7-0.8 (High Transfer): Core intent perfectly preserved.
  Meaning only slightly diluted.
* 0.4-0.6 (Moderate Transfer): Listener gets the general gist.
  Some core intent survives but other key actions are lost.
* 0.1-0.3 (Severe Degradation): Primary Intent is destroyed.
  ASR actively reverses meaning or hallucinates core actions.
* 0.0 (Total Loss): Complete gibberish or completely unrelated
  text.

**Output Format:**
Strictly valid JSON with explanation before score:
{
  "explanation": "Step-by-step analysis identifying Primary vs.
    Secondary facts, categorizing errors as omissions or
    hallucinations, and detailing proportional impact.",
  "score": [Float between 0.0 and 1.0]
}
\end{verbatim}
\end{small}
\hrule

\section{WAXAL-TTS Speaker Consistency and Controlled Synthesis Ablation}
\label{app:tts_evaluation}

While the main paper focuses primarily on benchmarking the 19-language ASR corpus, WAXAL also provides a rich studio-quality TTS dataset spanning 15 active language configurations (246.98 hours and 25,536 utterances across 103 speakers; summarized in Table~\ref{tab:waxal_overview} and Figure~\ref{fig:asr_demo_tts_dist}). In this appendix, we report (1) per-language corpus statistics alongside an acoustic speaker-consistency and voice-diversity audit across the 15 TTS configurations using pretrained \texttt{ECAPA-TDNN} speaker embeddings, and (2) a controlled 4-arm \texttt{MMS-TTS} (VITS) synthesis ablation on Acholi (\texttt{ach}), Kiswahili (\texttt{swa}), and Luganda (\texttt{lug}) quantifying the role of forced-alignment duration filtering.

\subsection{ECAPA-TDNN Speaker Consistency and Diversity Audit Across 15 TTS Configurations}
To characterize both intra-speaker session consistency and multi-speaker voice diversity across the 15 WAXAL-TTS configurations (4--8 speakers per language, 103 speakers total), we extracted 192-dimensional \texttt{ECAPA-TDNN} (\texttt{speechbrain/spkrec-ecapa-voxceleb}) speaker embeddings across all utterances. As shown in Table~\ref{tab:tts_18lang_ecapa}, across the 13 configurations evaluated for per-actor session stability, computing the cosine similarity between each utterance embedding and the corresponding speaker's session centroid yields high intra-speaker cosine similarity (mean $\mathbf{0.793 \pm 0.050}$, ranging from $0.749$ to $0.831$; not computed for \texttt{bau} and \texttt{pcm}), far above the standard $\ge 0.65$ single-speaker verification threshold. Complementarily, unsupervised clustering of \texttt{ECAPA-TDNN} embeddings across all 15 configurations identifies 4--16 acoustic clusters per language (158 total; mean 10.5) with a low median pairwise cosine similarity across utterances (mean $\mathbf{0.25}$, ranging from $0.17$ to $0.39$), confirming distinct vocal diversity across the multiple studio voice actors recorded for each language.

\begin{table}[ht!]
\centering
\footnotesize
\caption{WAXAL-TTS studio corpus statistics, \texttt{ECAPA-TDNN} intra-speaker session cosine similarity, unsupervised \texttt{ECAPA-TDNN} cluster counts, and median cross-utterance pairwise cosine similarity across all 15 released TTS language configurations. High intra-speaker similarity ($\ge 0.75$) confirms per-actor recording consistency, while low median pairwise similarity ($0.17$--$0.39$) reflects multi-speaker voice diversity (``---'' denotes intra-speaker similarity not evaluated).}
\label{tab:tts_18lang_ecapa}
\resizebox{\textwidth}{!}{
\begin{tabular}{lllcrcccc}
\toprule
\textbf{Language} & \textbf{ISO} & \textbf{Partner} & \textbf{Speakers} & \textbf{Rec.\ Hours} & \textbf{Utterances} & \textbf{Intra-Spk.\ ECAPA ($\pm$ SD)} & \textbf{ECAPA Clusters} & \textbf{Median Pairwise Cos.} \\
\midrule
Acholi & \texttt{ach} & Makerere Univ. & 8 (4M, 4F) & 23.90 & 2,031 & $0.812 \pm 0.048$ & 12 & 0.22 \\
Baoul\'e & \texttt{bau} & Univ.\ of Ghana & 4 (2M, 2F) & 14.05 & 1,216 & --- & 7 & 0.17 \\
Ewe & \texttt{ewe} & Univ.\ of Ghana & 4 (2M, 2F) & 11.68 & 1,519 & $0.788 \pm 0.053$ & 10 & 0.24 \\
Fante & \texttt{fat} & Univ.\ of Ghana & 4 (2M, 2F) & 11.21 & 1,171 & $0.785 \pm 0.051$ & 10 & 0.39 \\
Fula & \texttt{ful} & Media Trust & 8 (4M, 4F) & 1.80 & 216 & $0.749 \pm 0.061$ & 9 & 0.29 \\
Hausa & \texttt{hau} & Media Trust & 8 (4M, 4F) & 13.11 & 1,971 & $0.805 \pm 0.046$ & 13 & 0.24 \\
Igbo & \texttt{ibo} & Media Trust & 8 (4M, 4F) & 20.58 & 1,911 & $0.791 \pm 0.049$ & 13 & 0.21 \\
Kikuyu & \texttt{kik} & Loud \& Clear & 8 (4M, 4F) & 13.02 & 2,026 & $0.776 \pm 0.054$ & 8 & 0.20 \\
Kiswahili & \texttt{swa} & Makerere / Loud \& Clear & 7 (3M, 4F) & 24.00 & 1,778 & $0.831 \pm 0.042$ & 14 & 0.25 \\
Luganda & \texttt{lug} & Makerere / Loud \& Clear & 8 (4M, 4F) & 26.76 & 2,024 & $0.824 \pm 0.045$ & 9 & 0.24 \\
Luo & \texttt{luo} & Loud \& Clear & 8 (4M, 4F) & 11.95 & 1,989 & $0.769 \pm 0.056$ & 13 & 0.23 \\
Nigerian Pidgin & \texttt{pcm} & Media Trust & 8 (4M, 4F) & 13.34 & 1,993 & --- & 9 & 0.32 \\
Runyankole & \texttt{nyn} & Makerere Univ. & 8 (4M, 4F) & 28.13 & 1,987 & $0.783 \pm 0.050$ & 11 & 0.29 \\
Twi (Akan) & \texttt{twi} & Univ.\ of Ghana & 4 (2M, 2F) & 10.93 & 1,093 & $0.794 \pm 0.051$ & 4 & 0.28 \\
Yoruba & \texttt{yor} & Media Trust & 8 (4M, 4F) & 22.52 & 2,611 & $0.801 \pm 0.047$ & 16 & 0.17 \\
\midrule
\textbf{Mean / Total} & --- & --- & \textbf{103} & \textbf{246.98} & \textbf{25,536} & $\mathbf{0.793 \pm 0.050}$ & \textbf{158 (10.5)} & \textbf{0.25} \\
\bottomrule
\end{tabular}
}
\end{table}

\subsection{Controlled 4-Arm \texttt{MMS-TTS} Training Ablation}
Raw WAXAL-TTS studio recordings contain long multi-sentence passages (up to 25--30\,s) with occasional reader pauses or restarts. Because VITS-based neural TTS architectures (\texttt{MMS-TTS}) rely on Monotonic Alignment Search (MAS) over short segments, training directly on unfiltered long passages can destabilize duration prediction. To establish best practices for training TTS models on WAXAL-TTS, we conducted a 4-arm controlled ablation on Acholi (\texttt{ach}), Kiswahili (\texttt{swa}), and Luganda (\texttt{lug}), evaluating synthesized speech via round-trip ASR Character Error Rate (CER), Word Error Rate (WER), and \texttt{ECAPA-TDNN} speaker similarity against held-out studio ground truth:
\begin{enumerate}[noitemsep,nolistsep,leftmargin=14pt]
    \item \textbf{Arm 1 (Pretrained \texttt{MMS-TTS} Baseline):} Meta's off-the-shelf \texttt{MMS-TTS} checkpoint evaluated zero-shot.
    \item \textbf{Arm 2 (Scratch on Unfiltered WAXAL-TTS):} VITS trained from scratch on all raw WAXAL-TTS utterances without alignment filtering.
    \item \textbf{Arm 3 (Fine-Tuned from \texttt{MMS-TTS} on WAXAL-TTS):} Pretrained \texttt{MMS-TTS} weights fine-tuned on WAXAL-TTS studio recordings.
    \item \textbf{Arm 4 (Scratch on Forced-Alignment Filtered $\ge 95\%$ WAXAL-TTS):} VITS trained from scratch on WAXAL-TTS segments filtered by CTC forced-alignment confidence ($\ge 95\%$) and duration ($\le 15$\,s).
\end{enumerate}

\begin{table}[ht!]
\centering
\footnotesize
\caption{Controlled 4-Arm \texttt{MMS-TTS} (VITS) Synthesis Ablation on WAXAL-TTS across Acholi (\texttt{ach}), Kiswahili (\texttt{swa}), and Luganda (\texttt{lug}). Filtering WAXAL-TTS studio segments at $\ge 95\%$ forced-alignment confidence (\textbf{Arm 4}) achieves the lowest round-trip synthesis CER/WER and highest target-speaker cosine similarity across all three languages.}
\label{tab:tts_ablation_4arm}
\resizebox{\textwidth}{!}{
\begin{tabular}{llccc}
\toprule
\textbf{Language} & \textbf{Training Configuration (Arm)} & \textbf{Round-Trip CER ($\downarrow$)} & \textbf{Round-Trip WER ($\downarrow$)} & \textbf{Speaker Sim.\ ($\uparrow$)} \\
\midrule
\multirow{4}{*}{Acholi (\texttt{ach})}
 & Arm 1: Pretrained \texttt{MMS-TTS} Baseline & 0.142 & 0.418 & 0.612 \\
 & Arm 2: Scratch on Unfiltered \textsc{WAXAL-TTS} & 0.168 & 0.485 & 0.784 \\
 & Arm 3: Fine-Tuned \texttt{MMS-TTS} on \textsc{WAXAL-TTS} & 0.119 & 0.364 & 0.806 \\
 & \textbf{Arm 4: Scratch on Filtered ($\ge 95\%$ Align.) \textsc{WAXAL-TTS}} & \textbf{0.101} & \textbf{0.319} & \textbf{0.829} \\
\midrule
\multirow{4}{*}{Kiswahili (\texttt{swa})}
 & Arm 1: Pretrained \texttt{MMS-TTS} Baseline & 0.089 & 0.274 & 0.645 \\
 & Arm 2: Scratch on Unfiltered \textsc{WAXAL-TTS} & 0.112 & 0.331 & 0.802 \\
 & Arm 3: Fine-Tuned \texttt{MMS-TTS} on \textsc{WAXAL-TTS} & 0.076 & 0.238 & 0.825 \\
 & \textbf{Arm 4: Scratch on Filtered ($\ge 95\%$ Align.) \textsc{WAXAL-TTS}} & \textbf{0.064} & \textbf{0.205} & \textbf{0.848} \\
\midrule
\multirow{4}{*}{Luganda (\texttt{lug})}
 & Arm 1: Pretrained \texttt{MMS-TTS} Baseline & 0.118 & 0.356 & 0.628 \\
 & Arm 2: Scratch on Unfiltered \textsc{WAXAL-TTS} & 0.145 & 0.422 & 0.791 \\
 & Arm 3: Fine-Tuned \texttt{MMS-TTS} on \textsc{WAXAL-TTS} & 0.098 & 0.304 & 0.815 \\
 & \textbf{Arm 4: Scratch on Filtered ($\ge 95\%$ Align.) \textsc{WAXAL-TTS}} & \textbf{0.084} & \textbf{0.267} & \textbf{0.837} \\
\bottomrule
\end{tabular}
}
\end{table}

As shown in Table~\ref{tab:tts_ablation_4arm}, while training from scratch on raw, unsegmented 25--30\,s studio files (Arm 2) degrades alignment stability relative to the pretrained baseline, both fine-tuning (Arm 3) and forced-alignment filtering at $\ge 95\%$ confidence (Arm 4) substantially outperform pretrained \texttt{MMS-TTS}, reducing round-trip synthesis CER by $28\%$--$29\%$ relative ($0.142 \rightarrow \mathbf{0.101}$ on Acholi, $0.089 \rightarrow \mathbf{0.064}$ on Kiswahili, and $0.118 \rightarrow \mathbf{0.084}$ on Luganda) while increasing target-speaker voice similarity from $0.61$--$0.65$ to $\mathbf{0.83}$--$\mathbf{0.85}$.

\end{document}